\documentclass[a4paper,oneside,11pt]{article}

\usepackage[ansinew]{inputenc}

\usepackage[T1]{fontenc}
\usepackage{geometry}
\usepackage{amsmath}
\usepackage{amsthm}
\usepackage{amsfonts}
\usepackage{url}
\usepackage{graphicx}
\usepackage{multirow}
\usepackage{array}
\usepackage{lmodern}
\usepackage[english]{babel}

\begin{document}

\title{\textbf{Holographic vibrometry: Full field holographic vibrometry at ultimate limits} }

\author{Nicolas Verrier$^1$, Michael Atlan$^2$ and Michel Gross$^3$}

\date{}

\maketitle

\textit{$^1$Laboratoire Hubert Curien  UMR 5516 CNRS-Universit\'{e} Jean Monnet.
 18 Rue du Professeur Beno\^{\i}t Lauras 42000 Saint-Etienne}

 \textit{$^2$Institut Langevin.  UMR 7587- CNRS, U 979- INSERM, Universit\'{e} Pierre et Marie Curie (UPMC),
Universit\'{e} Paris 7.  Ecole Sup\'{e}rieure de Physique et de Chimie Industrielles (ESPCI) -1 rue Jussieu. 75005 Paris. France.}

\textit{$^3$ Laboratoire Charles Coulomb  - UMR 5221 CNRS-Universit\'{e} Montpellier II   Place Eug\`{e}ne Bataillon 34095 Montpellier
}
\abstract{ Heterodyne holography is an extremely versatile and powerful holographic technique that is able to fully control  the amplitude, phase and frequency of both the illumination and holographic reference beams. Applied to vibration analysis, this technique is able to detect  the signal of the carrier or the one of any vibration sideband. A complete analysis  of the vibration signal can thus be made, and  2D map of the vibration amplitude and phase   in all points of the surface of the vibrating object can be obtained.  Since the sensitivity is limited by shot noise, extremely low vibration amplitude (< 0.01 nm) can be studied.}

\paragraph{Citation} Verrier, N., Atlan, M. and Gross, M. (2015) Full Field Holographic Vibrometry at
Ultimate Limits, in New Techniques in Digital Holography (ed P. Picart), John Wiley \& Sons, Inc.,
Hoboken, NJ, USA. doi: 10.1002/9781119091745. ch7.

\section{Introduction}

Digital holography is a fast-growing research field
that has drawn increasing attention. The main advantage
of digital holography is that, contrary to holography
with photographic plates, the holograms
are recorded by a CCD, and the image of the object is digitally reconstructed
by a computer, avoiding photographic
processing \cite{schnars1994direct}.  To extract two quadratures of the holographic signal (i.e. to get the amplitude and the phase of the optical signal) two main optical configurations have been developed: off-axis and phase-shifting.

Off-axis holography is the oldest and
the simplest configuration adapted to digital
holography.  In that configuration, the reference or local oscillator
(LO) beam is angularly tilted with respect to the
object observation axis. It is then possible to record,
with a single hologram, the two quadratures of the
object complex field \cite{schnars1994direct}. However, the object field of view
is reduced, since one must avoid the overlapping of
the image with the conjugate image alias.
Phase-shifting digital holography \cite{yamaguchi1997phase} makes possible to get phase information on the whole camera area by recording several
images with a different phase for the reference (called here local oscillator or LO) beam.
It is then possible to obtain the two quadratures of
the field in an on-axis configuration even though the
conjugate image alias and the true image overlap, because
aliases can be removed by taking image differences.  In a typical phase-shifting holographic setup, the phase of the reference is shifted by moving a mirror with a PZT.

On the other hand, there is a big demand for full field vibration measurements, in particular in industry. Different
holographic techniques are able to image and analyze such vibrations. Double pulse holography
\cite{pedrini1995digital,pedrini1997digital} records a double-exposure hologram with time separation in the 1...1000 $\mu$s range, and
measures the instantaneous velocity of a vibrating object from the phase difference. The method
requires a quite costly double pulse ruby laser system, whose repetition rate is low. Multi pulse
holography \cite{pedrini1998transient} is able to analyse transient motions, but the setup is still heavier (4 pulses laser,
three cameras).

The development of fast CMOS camera makes possible to analyze vibration efficiently by
triggering the camera on the motion in order to record a sequence of holograms that allows to
track the vibration of the object as a function of the time \cite{pedrini2006high,fu2007vibration}. The  analysis of the
motion can be done by phase difference or by Fourier analysis in the time domain. The method
requires a CMOS camera, which can be costly. It is also limited to low frequency vibrations,
since a complete analysis of the motion requires a camera frame rate higher than the vibration
frequency, because the bandwidth $\textrm{BW}$ of the holographic signal, which is sampled at the camera at angular frequency $\omega_{CCD}$ must be lower than corresponding Nyquist-Shannon limit: $BW < \frac{1}{2}~\omega_{CCD}$.

For a periodic vibration motion, the bandwidth $\textrm{BW}$ is close to zero.  Measurements can thus be done with much slower cameras.
Powell and Stetson \cite{powell1965interferometric} have shown for example that an harmonically vibrating
object yields  alternate dark and bright fringes, whose analysis yields informations on the vibration motion.
Picard et al. \cite{picart2003time} has simplified the processing of
the data by performing time averaged holography with a digital CCD camera.
Time averaged
holography has no limit in vibration frequency and do not involve costly laser system, nor an
expensive CMOS fast camera. Although the time-averaging method gives a way to determine the amplitude
of vibration \cite{picart2005some} quantitative measurement remain quite difficult for low and high vibration
amplitudes.

Heterodyne holography \cite{le2000numerical,le2001synthetic} is a variant of phase shifting holography, in which the frequency, phase and amplitude of both reference and signal signal beam are  controlled by acousto optic modulators (AOM).  Heterodyne holography is thus extremely versatile.
By shifting the frequency  $\omega_{LO}$ of the local oscillator beam with respect to the  frequency  $\omega_0$ of illumination, it is for example possible  to detect the holographic signal at a frequency $\omega$ different  than illumination $\omega_0$. This ability will be extremely useful to analyze vibration, since  heterodyne holography can detect selectively the signal that is scattered by the object on a vibration sideband of frequency   $\omega_m=\omega_0+ m \omega_A$, where $\omega_A$ is the vibration frequency and  $m$ and integer index.

In this chapter we will first present in section  \ref{section_Heterodyne holography}
heterodyne holography and its advantages in section \ref{section_accurate_phase} and \ref{section_shot_noise}. Then in section  \ref{section_vibrometry}, we will apply  heterodyne holography to vibration analysis. We will show in section \ref{section_Selective detection of the sideband components}, how heterodyne holography  can be used to detect the vibration sidebands, in section \ref{section_strobe_detection} how this sideband holography can be combined with stroboscopic techniques to record instantaneous velocity maps of motion, and in sections  \ref{Section_large_vibration} and \ref{section_small_vibration} how it can retrieve both small and large vibration amplitudes.


\section{Heterodyne holography}\label{section_Heterodyne holography}


\begin{figure}
\begin{center}
  \includegraphics[width=9 cm]{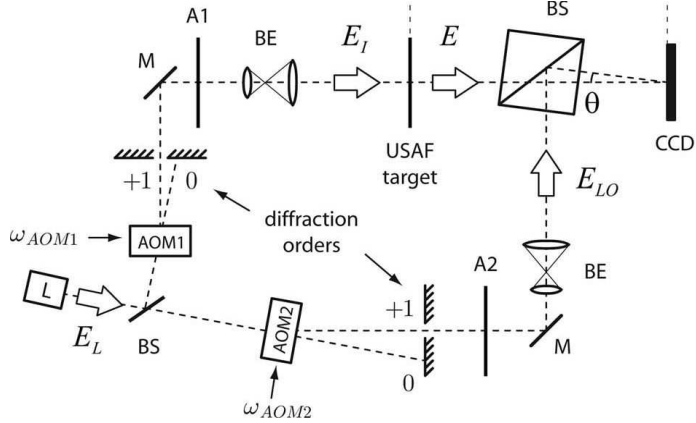}\\
 \caption{Typical heterodyne holography setup. L: main
laser; BS: beam splitter; AOM1, AOM2: acousto-optic
modulators; BE: beam expander; M, mirror; A1, A2: light
attenuators; USAF, transmission USAF 1951 test pattern;
CCD:  camera; $E_L$, $E_I$, $E_{LO}$, and $E$: laser, illumination,
reference (i.e. local oscillator LO) and object fields; $\omega_{AOM1/2}$: driving frequencies of the
acousto optics modulators AOM1 and AOM2; $\theta$: off-axis angular tilt.}\label{Fig_fig_heterodyne_setup}
\end{center}
\end{figure}

Let us first describe heterodyne holography. A example of setup is shown on Fig. \ref{Fig_fig_heterodyne_setup}. The object is an U.S. Air Force (USAF) resolution target whose hologram is recorded in transmission geometry (the target is back illuminated). The camera records the interference of the signal field ${\cal E}(t)$ of optical angular frequency $\omega_0$, with the reference (or local oscillator field) ${\cal E}_{LO}(t)$ of optical frequency
\begin{eqnarray}
 \omega_{LO}&=&\omega_0- \Delta \omega
\end{eqnarray}
 where $\Delta \omega$ is the frequency shift. Let us introduce the slowly varying complex fields $E$ and $E_{LO}$.
\begin{eqnarray}
 {\cal{E}} (t)& = &E e^{j\omega_0 t} + E^* e^{-j\omega_0 t}\\
\nonumber {\cal{E}}_{LO} (t) &= &E_{LO}  e^{j\omega_{LO}  t} + E_{LO} ^* e^{-j\omega_{LO}  t}
\end{eqnarray}
where $E^*$ and $E_{LO} ^*$ are the complex conjugates of $E$ and $E_{LO}$, and $j$ is the imaginary unit. The camera signal $I$ is proportional to the intensity intensity of the total field $ |{\cal{E}} (t)+{\cal{E}}_{LO} (t)|^2$. We have thus:
\begin{eqnarray}\label{Eq_I}
   I&=&\left|E_{LO}~ e^{j\omega_{LO}t}~+ ~E~ e^{j\omega_0 t}\right|^2 \\
 \nonumber  &=& \left|E_{LO} \right|^2 + \left|E \right|^2 + E_{LO}^* E~e^{+j\Delta \omega t} +  E_{LO} E^*~e^{-j\Delta \omega t}
\end{eqnarray}
Since the camera signal is slowly varying, we have neglected in Eq. \ref{Eq_I} the fast varying terms  (which evolve at frequency $\simeq 2\omega_0$). Moreover, to simplify the present discussion, we  have not considered  the spatial variations of $I$, $E$ and $E_{LO}$ with $x$ and $y$, in particular the spatial variations that are related to the off axis tilt angle $\theta$ of Fig. \ref{Fig_fig_heterodyne_setup}.
In  equation \ref{Eq_I},   $E_{LO}^* E~e^{-j\Delta \omega t}$ is the +1 grating order term, that contains the useful information (since this term  is proportional to $E$). The others terms: $\left|E_{LO} \right|^2 + \left|E \right|^2$ (zero grating order term),    and  $E_{LO} E^*~e^{+j\Delta \omega t}$ ( -1 grating order term or twin image term) are unwanted terms that must be cancelled. To filter off these terms, and to select the wanted +1 grating order signal,   4-phase detection is made. The AOMs driving angular frequencies  $\omega_{AOM1}$ and  $\omega_{AOM2}$ are tuned to have:
\begin{eqnarray}\label{Eq_4_phases}
  \Delta \omega =  \omega_{AOM2}-  \omega_{AOM1} =\omega_{CCD}/4
\end{eqnarray}
where $\omega_{CCD}$ is is the angular frequency of the camera frame rate. Lets us consider 4 successive camera frames: $I_0$, $I_1$ ...$I_3$  that are recorded at times $t=0$, $T$ ... $3T$ with $T = 2 \pi/( \omega_{CCD})$.
For these  frames, the  phase factor  $e^{+j\Delta \omega t}$ is equal to $1$, $j$, -1 and $-j$. We thus get:
\begin{eqnarray}\label{Eq_4_phases_I}
  I_0&=&    \left|E_{LO} \right|^2 + \left|E \right|^2 + ~E_{LO}^* E + ~ E_{LO} E^*\\
\nonumber  I_1&=&    \left|E_{LO} \right|^2 + \left|E \right|^2 +j~ E_{LO}^* E - j ~E_{LO} E^*\\
\nonumber  I_2&=&    \left|E_{LO} \right|^2 + \left|E \right|^2 -~E_{LO}^* E  - ~E_{LO} E^*\\
\nonumber  I_3&=&    \left|E_{LO} \right|^2 + \left|E \right|^2 - j ~E_{LO}^* E  + j~E_{LO} E^*
\end{eqnarray}
By linear combination of $4$ frames, we get the 4-phase hologram $H$ that  obeys the demodulation equation:
\begin{eqnarray}\label{Eq_4_phase_Hs}
 H&=& (I_0-I_2)+j(I_1-I_3)\\
 \nonumber &=& 4~E_{LO}^* E
\end{eqnarray}
As wanted, the 4-phase demodulation equation yields a quantity $H$ that proportional to the signal complex field $E$. Note that the coefficient $4 E_{LO}$ is supposed to be known and do not depends on the object.

Heterodyne holography exhibits several advantages with respect to other holographic
techniques:
\begin{enumerate}
  \item The phase shift is very accurate;
  \item The holographic detection is shot noise limited;
  \item Since the holographic  detection is  made somewhere near $\omega_{LO}$ (depending on the demodulation equation that is chosen), it is possible to perform the holographic detection  with any frequency shift with respect to the object illumination angular frequency $\omega_0$.
\end{enumerate}
The first advantage will be discussed in section \ref{section_accurate_phase},
the second one in section \ref{section_shot_noise}, while the third one, which  is the heart of the sideband holographic technique used to analyse vibration, will be discussed in
section \ref{section_Selective detection of the sideband components}.

\subsection{Accurate phase shift  and holographic detection bandwidth}\label{section_accurate_phase}

\begin{figure}[h]
\begin{center}
    \includegraphics[width=8 cm]{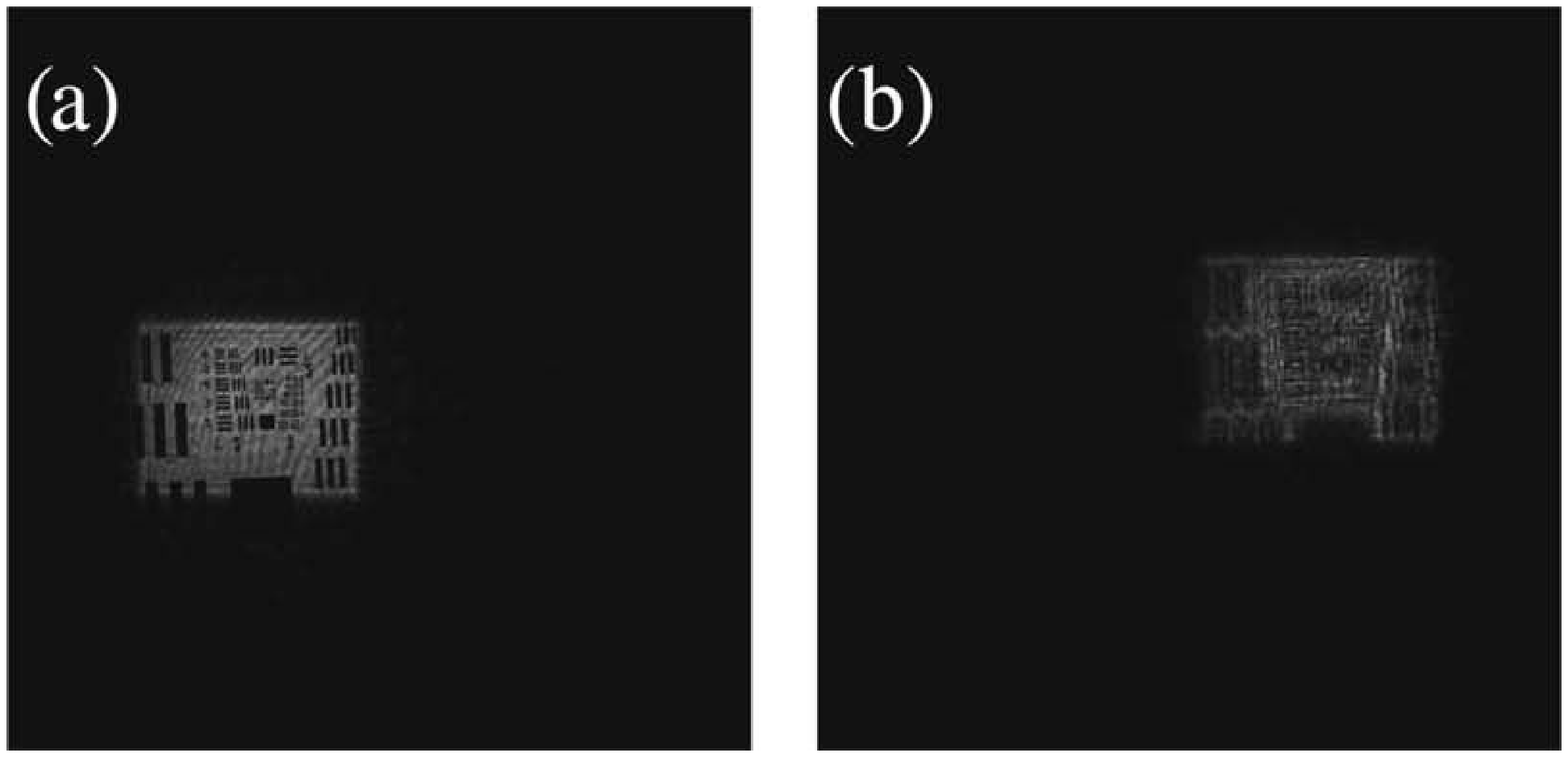}
  \includegraphics[width= 8 cm]{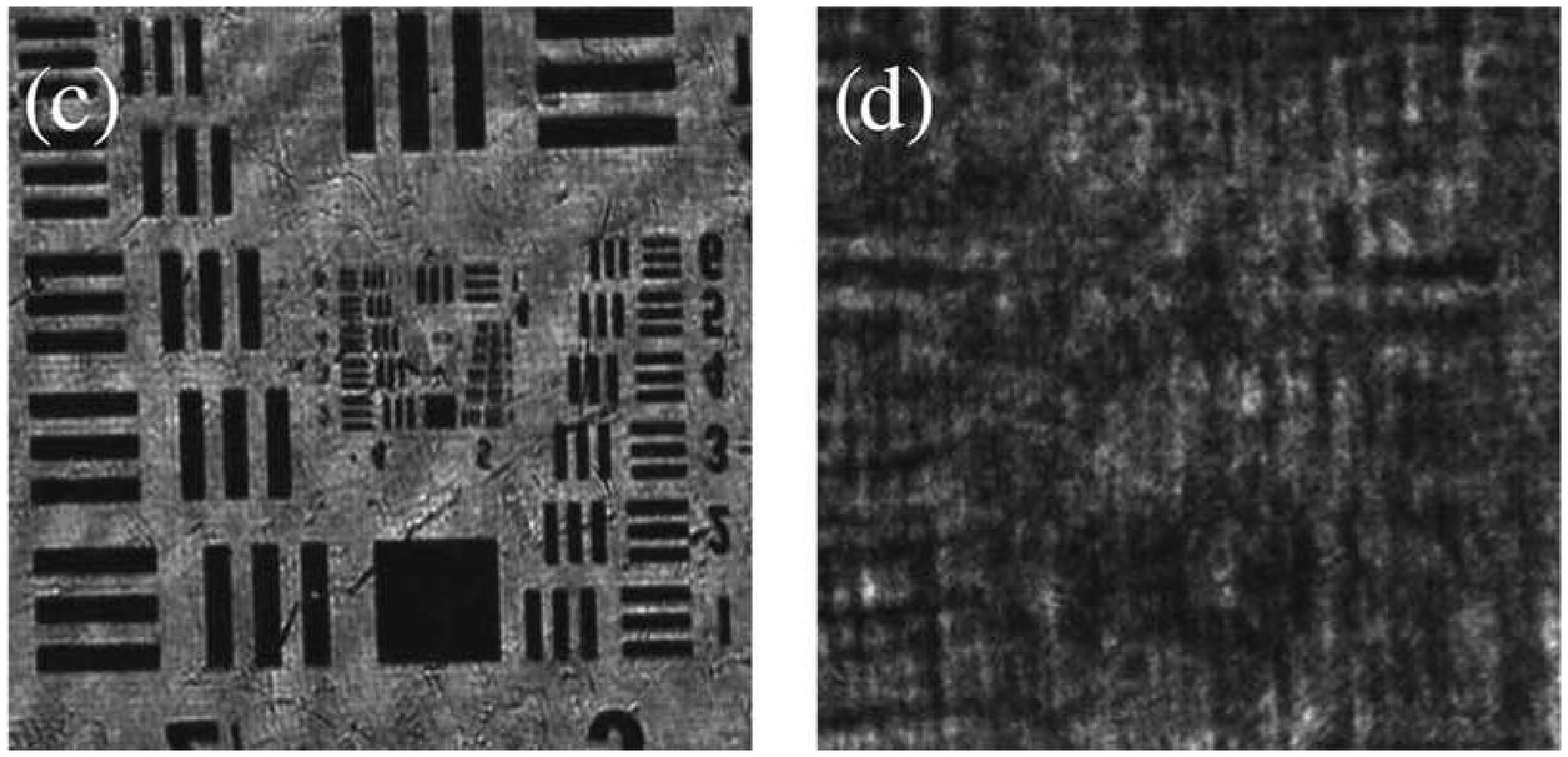}
 \caption{ USAF target reconstructed images (a,b) displayed with linear grey scale for the reconstructed  field intensity  $|E|^2$ with    $ \omega_0- \omega_{LO} = \omega_{CCD}/4=+2.5$ Hz (a) and   $ \omega_0- \omega_{LO} = - \omega_{CCD}/4= -2.5$ Hz (b).  Zooms  (c,d)  of the bright zones of the (a) and (b) images. Images (a,b) and zoom (c,d)  correspond to the $m=+1$ (a,c) and $m=-1$ (b,d) grating orders. }\label{Fig_fig_usaf_accurate_phase_abcd}
\end{center}
\end{figure}

In the  typical heterodyne holography setup of Fig.\ref{Fig_fig_heterodyne_setup}, the signals that drive the  AOMs  are generated by frequency synthesizers, phase-locked with a common 10 MHz clock.. The phase shift  $\Delta \varphi= \Delta \omega T$ (that is equal to  $\pi/2$ in the 4 phase demodulation case) can thus be adjusted with quartz accuracy.
%

To illustrate this accurate phase shift,  holograms of a USAF target have been recorded  with the Fig.\ref{Fig_fig_heterodyne_setup} setup, while sweeping the LO frequency $\omega_{LO}$. Figure \ref{Fig_fig_usaf_accurate_phase_abcd} shows  USAF reconstructed images that are obtained for different values of the frequency shift $ \omega_0-\omega_{LO}$  \cite{atlan2007accurate}.
\begin{enumerate}
  \item For $\omega_0- \omega_{LO} = \omega_{CCD}/4=+2.5$ Hz, the image that is reconstructed is sharp and  correspond to the +1 grating order.  On the other hand, the -1 grating order signal is very low; the magnitude of the -1 parasitic twin image image is negligible in front of the +1 contribution. Image and zoom of image are seen on Fig.\ref{Fig_fig_usaf_accurate_phase_abcd} (a) and (c).
  \item For $\omega_0- \omega_{LO}  = -\omega_{CCD}/4= -2.5$ Hz, the image that is reconstructed is blurred and  corresponds to the -1 grating order twin image. The magnitude of the +1 grating order image is negligible in front of its -1 counterpart. Image and zoom of image are seen on Fig.\ref{Fig_fig_usaf_accurate_phase_abcd} (b) and (d).
\end{enumerate}

\begin{figure}
\begin{center}
  \includegraphics[width=12 cm]{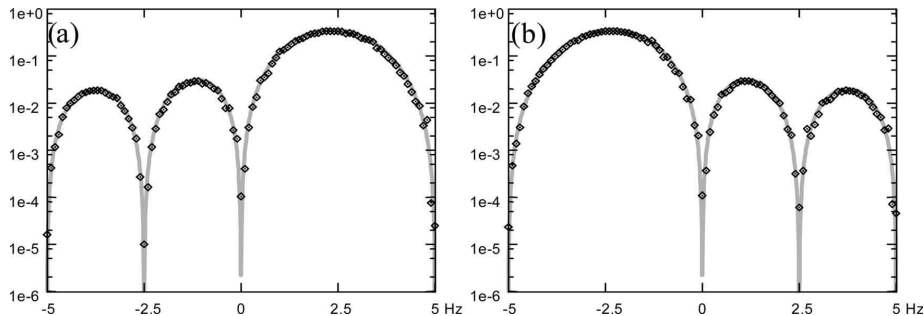}
 \caption{ Field energy  of the +1 (a) and -1 (b) grating order. The vertical axis is in logarithmic scale arbitrary units. The points correspond to the experimental data obtained by sweeping the LO frequency  with   $(\omega_0-\omega_{LO}) =-5$ to +5 Hz with 0.1 Hz increments. The solid gray curves are  $+U_{+}(\omega-\omega_{LO})$ (a) and $+U_{-}(\omega-\omega_{LO})$ (b) given by  Eq.\ref{Eq_total_versus_delta_omega} in the same frequency range:   $(\omega-\omega_{LO}) =-5$ to +5 Hz. }\label{Fig_fig_accurate_phase_curves}
\end{center}
\end{figure}

We have measured the total field energy $U_{\pm}$ in both +1 and -1 grating order reconstructed images:
\begin{eqnarray}\label{Eq_total_energy_definition}
  U_{\pm} &=&   \sum_{x,y}\left|E(x,y) \right|^2
\end{eqnarray}
where $\sum_{x,y}$ is the sum over  the pixels of the reconstructed field image $E(x,y)$ corresponding to the  $\pm 1$ grating order zones.
As seen on Fig. \ref{Fig_fig_accurate_phase_curves}, the energy of the signal that is measured in the +1 grating order (i.e. $U_+$)    is maximum for  $\omega_0-\omega_{LO}=\omega_{CCD}/4 = +2.5$ Hz, and null  $\omega_0 -\omega_{LO}=-2.5$ Hz. Similarly,  $U_-$ is maximum for  $ \omega_0 -\omega_{LO}=-2.5$ Hz, and null for $\omega_0 -\omega_{LO}=+2.5$ Hz.  By adjusting $\omega_0 -\omega_{LO}$, it is then possible to select the grating order that is detected.

For 4 phase detection with a local oscillator of frequency $\omega_{LO}$, the total energy  $U_{\pm}=|E|^2$ detected at frequency $\omega$ in the $\pm 1$ grating order can be easily calculated   \cite{verpillat2010digital}:
\begin{eqnarray}\label{Eq_total_versus_delta_omega}
   U_{\pm}(\omega-\omega_{LO}) &=& \left| ~\frac{1}{4T'}\sum_{n=0}^{n=3} (\pm j)^n \int_{t=nT}^{nT+T'}~dt~ e^{ \pm j ( \omega -\omega_{LO}) t} \right|^2\\
 \nonumber  &=& \left|\frac{ \textrm{sinc}(\pi (\omega-\omega_{LO}) T')}{4} \sum_{n=0}^{n=3} j^n
    e^{j  n (\omega-\omega_{LO})T} \right|^2
\end{eqnarray}
where $ \textrm{sinc} (x) = \sin x/x$. In equation \ref{Eq_total_versus_delta_omega}, $T$ is the frame period, and  $T'$ the exposure time. The coefficient   $1/4T'$ is thus a normalization factor. On the other hand,
the factors $(\pm j)^n$ corresponds to the coefficients of  the   demodulation equation (see Eq. \ref{Eq_4_phase_Hs}), and   $ e^{\pm j (\omega-\omega_{LO})  t}$ is the interference instantaneous phase factor, which must be integrated  over the exposure time from  $t=nT$ to $t=nT+T'$.

In the  experiment of Fig.\ref{Fig_fig_accurate_phase_curves}, we have $T=T'=2\pi/\omega_{CCD}=100 $ ms. We have plotted the field energy  $U_{\pm}$ given by Eq.\ref{Eq_total_versus_delta_omega} as a function of $ \omega- \omega_{LO}$ on Fig. \ref{Fig_fig_accurate_phase_curves} (solid grey lines). As seen, experiment agrees  with the theoretical curve of Eq. \ref{Eq_total_versus_delta_omega}. The shape of the curves  represents here the frequency response spectrum of the holographic device considered as a detector. For the +1 grating order, detection is centered at frequency $ \omega = \omega_{LO} + \omega_{CCD} / 4$. The measurement bandwidth  BW is  2.5 Hz. It is equal to the inverse of the measurement time of 4 frames i.e. $\textrm{BW}= 1 / 4T$.  It illustrates the coherent character (in time) of holographic detection.

\begin{figure}
\begin{center}
  \includegraphics[width=5.5 cm]{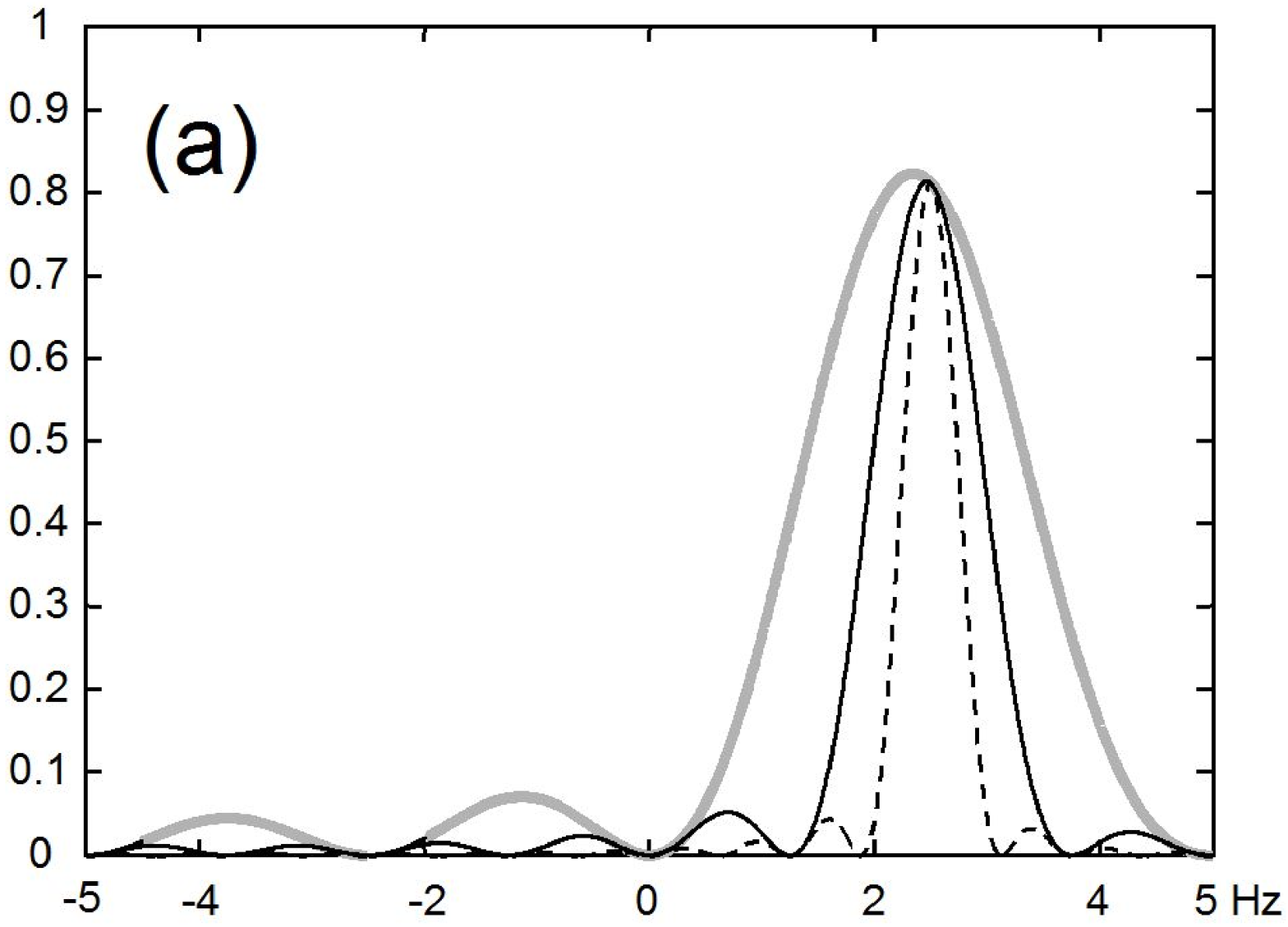}
  \includegraphics[width=5.5 cm]{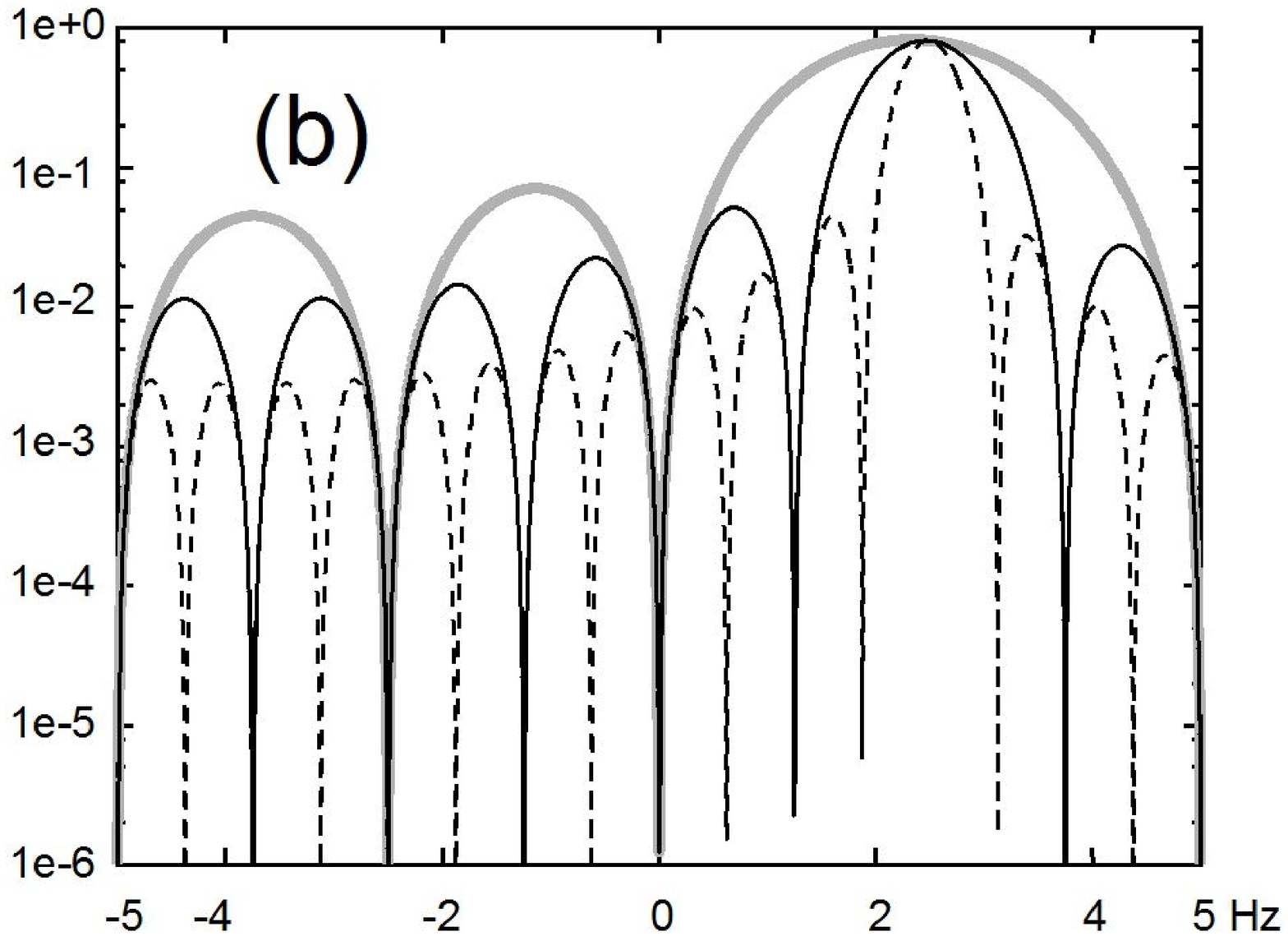}
 \caption{ Theoretical field energy  $U_{+}(\omega-\omega_{LO})$ given by Eq.\ref{Eq_total_versus_delta_omega}    plotted in linear (a) and  logarithmic scale (b) as a function of $( \omega-\omega_{LO})$ in Hz Units for the +1  grating order. The number of frame is $n_{max}=4$ (solid grey line), $n_{max}=8$ (solid black line) and $n_{max}=16$ (dashed black line).
 }\label{Fig_fig_U_curves}
\end{center}
\end{figure}

It is possible to increase the selectivity of the holographic coherent detection by increasing the measurement time, i.e. by increasing the number of frames $n_{max}$ used for demodulation. In that case, Eq. \ref{Eq_4_phase_Hs} and Eq.\ref{Eq_total_versus_delta_omega}  must be replaced by similar equations involving   $n_{max}$ frames in place of 4. We have:
\begin{equation}\label{EQ_H_n_max}
    H= \sum_{n=0}^{n_{max}-1} j^n ~I_n
\end{equation}
and
\begin{eqnarray}\label{Eq_total_versus_delta_omega_nmax}
   U_{\pm}(\omega-\omega_{LO}) &=& \left| ~\frac{1}{n_{max}T}\sum_{n=0}^{n_{max}-1} (\pm j)^n \int_{t=nT}^{nT+T'}~dt~ e^{\pm j (\omega-\omega_{LO}) t} \right|^2\\
    \nonumber  &=& \left|\frac{ \textrm{sinc}(\pi (\omega-\omega_{LO}) T')}{n_{max}} \sum_{n=0}^{n_{max}-1} j^n
    e^{j n (\omega-\omega_{LO})T} \right|^2
\end{eqnarray}
Figure  \ref{Fig_fig_U_curves} plots  the  detection frequency spectrum $U_{+}$   for $n_{max}=4$, 8 and 16.    As seen, the detection bandwidth BW  decreases with $n_{max}$. It is equal to:
\begin{equation}\label{Eq_BW}
    \textrm{BW} = \frac{1}{n_{max} T}
\end{equation}
The noise, uniformly distributed in frequency, is expected to decrease accordingly.  We will show in section \ref{section_shot_noise}, that very high detection sensitivity can  be obtained by holographic detection with large number of frames $n_{max}$.

\subsection{Shot noise holographic detection}\label{section_shot_noise}

Because the holographic signal results in the interference of the object
signal complex field $E$ with a reference (LO) complex
field $E_{LO}$ whose amplitude can be much larger (i.e.,
$E_{LO} \gg E$),  the holographic detection benefits from ''heterodyne'' or ''holographic'' gain (i.e.,  $|E E^*_{LO}| \gg |E|^2$), and is thus well suited for  detection  of  weak signal fields $E$.  Holographic detection can reach the theoretical limit of noise which corresponds to a noise equivalent signal of 1 photo electron per pixel during the total measurement time~\cite{gross2007digital,gross2008noise,verpillat2010digital,lesaffre2012noise}.

\begin{figure}
\begin{center}
  \includegraphics[width=11 cm]{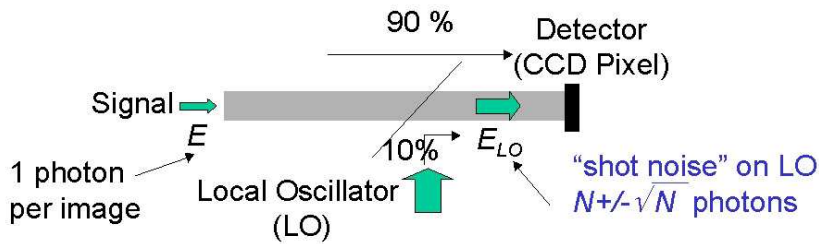}
 \caption{Heterodyne detection of 1 photon per pixel of signal with N photon
local oscillator.
 }\label{Fig_fig2_shot_noise}
\end{center}
\end{figure}

To illustrate this point, let us consider the interference of a weak signal  of about 1 photo-electron per pixel and per camera frame, with a large local oscillator reference signal  of $N=10^4$ photo-electrons: see Fig. \ref{Fig_fig2_shot_noise}. In this example, the signal intensity is $|E|^2=1$, while the holographic interference term $|E E^*_{LO}|=100$ is much larger. The detected signal is  $|E_{LO}+E|^2 \simeq 10^4$ photo-electrons.

Because of the quantum nature of the process involved in digital holography (laser emission, photodetection...), the detected signal in photo electron Units is random Gaussian integer number, whose average value is $ N=10^4$, and whose standard deviation is $ \sqrt{N}=100$. These  fluctuations of the number of photo electrons are called shot noise. Here, with 1 photo electron of signal, the heterodyne signal is  $100$ photo electrons and the noise  100 too. The noise equivalent signal (for the energy $|E|^2$) is thus 1 photo electron per pixel and per frame.

Let us now study how shot noise varies with the number of frames  $n_{max}$ used for  detection.  As in any detection process, the noise in energy is proportional to the measurement time and to the detection bandwidth BW. Since  the measurement time is $n_{max} T$, and since  the detection bandwidth is $\textrm{BW}=1/n_{max} T$ (see Eq. \ref{Eq_BW} ),  the shot noise in energy do not depends on $n_{max}$. The shot noise equivalent signal is thus the same than for one frames. It remains equal to   1 photo electron per pixel whatever the number  of frames $n_{max}$ and measurement time $n_{max}T$ are.

Let us now discuss the ability to reach this  shot noise optimal sensitivity  in
real life holographic experiments. Since we consider implicitly
a very weak signal, the noises that must be
considered are
\begin{enumerate}
  \item the read noise and dark current of the  camera,
  \item the quantization noise of the camera A/D converter,
  \item the technical noise of the LO beam,
  \item and the LO beam shot noise, which yield the theoretical
noise limit.
\end{enumerate}
For a typical camera, the full well capacity is $2 ~10^4$ photo electrons, and a good practice is to work with $N=10^4$ photo electrons for the local oscillator. Shot noise on the camera signal (100 photo electrons) is thus much larger than the camera read noise (1 to 20 photo electrons) and the camera dark current (a few photo electrons per second). If the camera is 12 bit, the  full well capacity   corresponds $\sim 2^{12}$ digital count (DC). As a results, the quantization noise ($\sim 7 $ photo electrons) can be neglected too.

\begin{figure}
\begin{center}
  \includegraphics[width=8 cm]{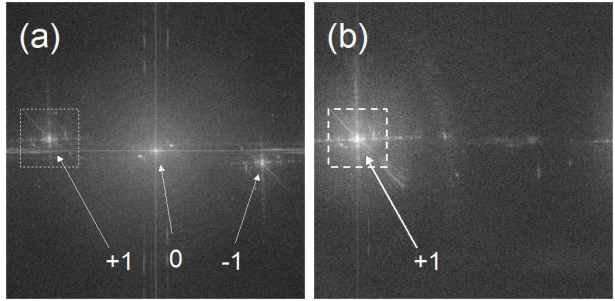}
 \caption{Fourier space hologram intensity (i.e. $|\tilde H(k_x,k_y)^2$) displayed in arbitrary log scale for one phase ($H=I_0$) and 4 phases ($H=(I_0-I_2)+j(I_1-I_3)$ detection.
 }\label{Fig_fig_usaf_1_4phases}
\end{center}
\end{figure}

Let us discuss on the technical noise of the LO beam, and on the way to filter it off.
\begin{enumerate}
  \item
We have displayed,   on Fig. \ref{Fig_fig_usaf_1_4phases}, examples of Fourier space hologram $\tilde H(k_x,k_y) =\textrm{FFT}( ~H(x,y))$ (where FFT is the Fourier transform operator). The signal fields yields the interference term $E E^*_{LO}$ that is located in the left hand side of the Fourier space images of Fig. \ref{Fig_fig_usaf_1_4phases} (a) and (b). It corresponds to the grating order +1.  The local oscillator signal  $|E_{LO}|^2$ is located in the center of the Fourier space. It is visible on Fig. \ref{Fig_fig_usaf_1_4phases} (a) that is obtained with 1 phase hologram. As seen, the LO signal are  separated in the Fourier space  because of the off axis configuration.
It is then possible to filter off the parasitic LO signal by a proper spatial filtering in the Fourier space as shown by  \cite{cuche2000spatial}.
As noticed, this operation filters-off both  the LO signal and the LO technical noise.
  \item On the other hand, because of phase shifting, the  signal and LO   fields  $E$ and $E_{LO}$  evolve at two different time frequencies  $\omega$ and $\omega_{LO}$ with $\omega=\omega_{LO} + \omega_{CCD}/4$. It is then possible to filter off the LO signal in time. This is done by the 4 phases demodulation process since the LO term $|E_{LO}|^2$ vanish in Eq. \ref{Eq_4_phase_Hs} and \ref{Eq_total_versus_delta_omega_nmax}. This filtering in time is illustrated by Fig. \ref{Fig_fig_usaf_1_4phases}, since the 0 grating order signal that is large for single phase off-axis holography  (Fig. \ref{Fig_fig_usaf_1_4phases} (a)), roughly vanishes in the four phase case (Fig. \ref{Fig_fig_usaf_1_4phases} (b)).
Time filtering is also illustrated by  Fig.\ref{Fig_fig_accurate_phase_curves} and Fig.\ref{Fig_fig_U_curves} and Eq. \ref{Eq_4_phase_Hs} and \ref{Eq_total_versus_delta_omega_nmax}, since the detected energy  $U_{+/-}$ vanishes for  $\omega- \omega_{LO} =0 $ i.e. for detection at local oscillator frequency $\omega_{LO}$. Here again, one filters off both  the LO signal and the LO technical noise.
\end{enumerate}
By combining off axis and phase shifting, it is then possible apply to a double filtering (in space and time) that filter off the LO technical noise very effectively. One can gets then the shot noise ultimate sensitivity that corresponds to a noise equivalent signal of  1 photon per pixel whatever the measurement times is.

\begin{figure}
\begin{center}
\includegraphics[width=12 cm]{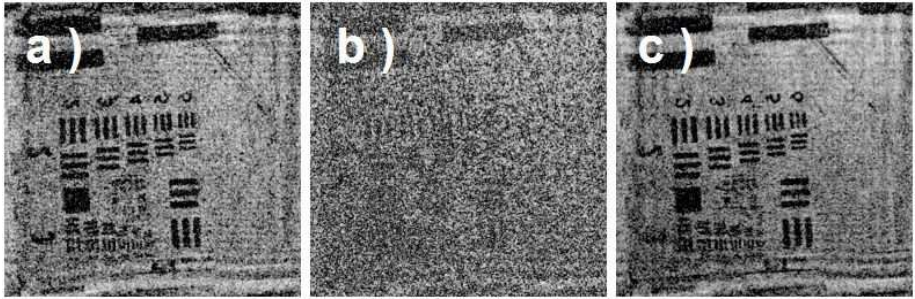}
 \caption{Reconstructed image of a USAF target in dim light. Hologram are recorded by 4 phases heterodyne holography with $T=100$ ms, and $n_{max}=600$ (a,b) and $n_{max}=6000$  (c). The coherent measurement time $n_{max} T $ is thus 1 minute (a,b) and 10 minutes (c). Illumination is adjusted so that the USAF signal integrated is   1 photon per pixel over the whole measurement time $n_{max}T$  in (a). Holograms (b) and (c) are recorded by adding a neutral density filter D=1.0 on illumination. The USAF signal is thus 0.1 photon per pixel in (b) and  1 photon per pixel in (c). Display is made with arbitrary linear scale for intensity.}\label{Fig_fig_usaf_10mn}
\end{center}
\end{figure}

Figure \ref{Fig_fig_usaf_10mn} illustrates  heterodyne holography ability of imaging an object
(here an USAF target) in dim light with shot noise limited sensitivity. In figure \ref{Fig_fig_usaf_10mn} (a), the measurement time is 1 minute and the illumination power is adjusted such a way the USAF signal is about 1 photon per pixel in 1 minute. The visual quality of USAF  image is quite good. We can say that  SNR is about 1.  In figure \ref{Fig_fig_usaf_10mn} (b), the  measurement time is the same, but illumination is divided by 10 by using a neutral  filter of density $D=1.0$. SNR is low, and one cannot see the grooves of the USAF target. Figure \ref{Fig_fig_usaf_10mn} (c) is obtained with the same illumination level that Fig. \ref{Fig_fig_usaf_10mn} (b) (i.e. with neutral density filter) but the measurement time is multiplied by 10 (i.e. 10 minutes in place of 1 minute). The visual quality of reconstructed image is the same than for Fig. \ref{Fig_fig_usaf_10mn} (a) with  SNR about 1. This experiment shows that the reconstructed image quality depends on the total amount of signal, and do not depends on the  time needed to get that amount of signal: to get $\textrm{SNR}\sim 1$ one needs about 1 photon per pixel, for any measurement time.


\section{Holographic vibrometry}\label{section_vibrometry}

Lets us now  apply heterodyne holography
to vibration analysis. 

\subsection{Optical signal scattered by a vibrating object}

\begin{figure}
\begin{center}
\includegraphics[width=9 cm]{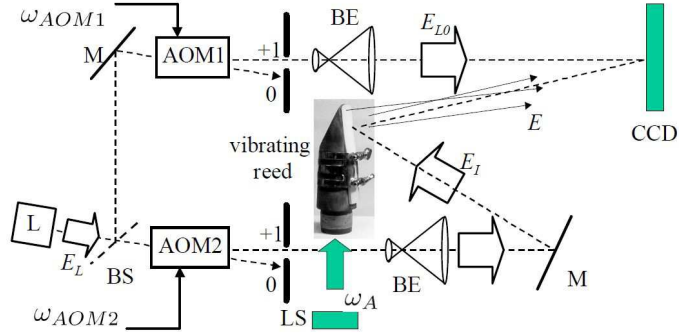}
 \caption{Heterodyne holography setup applied to analyse vibration of a clarinet reed. L: main laser; AOM1, AOM2: acousto-optic modulators; M: mirror; BS:
beam splitter; BE: beam expander; CCD:  camera; LS: loud-speaker exiting the vibrating  clarinet
reed at frequency $\omega_A/2\pi$.}\label{Fig_fig_setup_reed}
\end{center}
\end{figure}

Consider a point of the  objet  (for example a clarinet reed) that is studied by heterodyne holography (see Fig. \ref{Fig_fig_setup_reed}), vibrating  at frequency
$\omega_A$  with amplitude $z_{max}$. The displacement $z$ along the out of plane direction is
\begin{equation}\label{Eq_z_t}
  z(t) = z_{max} \sin{\omega_A t}
\end{equation}
In backscattering geometry, this corresponds to a
phase modulation $\varphi(t)$ of the signal:
\begin{eqnarray}\label{Eq_varphi_t}
 \varphi(t) &=& 4\pi z(t)/\lambda\\
  \nonumber  &=&\Phi \sin{\omega_A t}
\end{eqnarray}
where $\lambda$ is the optical wavelength and  $\Phi$ the amplitude of the phase modulation of the signal  at angular frequency $\omega_A$:
\begin{eqnarray}\label{Eq_Phi}
\Phi&=&4\pi z_{max}/\lambda
\end{eqnarray}
Let us define the slowly varying complex  amplitude $E(t)$ of the field ${\cal E}(t)$ scattered by the vibrating object. We have:
\begin{eqnarray}
{\cal E}(t)&=& E(t) e^{j\omega_0 t} + \textrm{c.c.}
\end{eqnarray}
Because of the Jacobi-Anger expansion, we get:
\begin{eqnarray}\label{Eq_cal_E_sum_over_harmonic}
\nonumber E(t)&=&E ~e^{j\varphi(t)}=E ~e^{j \Phi \sin{\omega_A t}  }\\
\nonumber &=& E ~\sum_m J_m(\Phi)~e^{j n\omega_A t}
\end{eqnarray}
where $E$ is the complex amplitude without vibration, and $J_m$
 the mth-order Bessel function of the first kind, with
$J_{-m}(z)=-1^m J_m(z)$ for integer $ m$ and real $z$.
The scattered
field ${\cal E}(t)$ is then the sum of the carrier and sideband field components ${\cal E}_m(t)$ of frequency $\omega_m$, where $m$ is the sideband index with:
\begin{eqnarray}\label{Eq_cal_E_m}
 {\cal E}(t) &=& \sum_{m=-\infty}^{+\infty}{\cal E}_m(t)\\
\nonumber   {\cal E}_m(t) &=& E_m e^{j\omega_m t} +  E_m^* e^{-j\omega_m t}\\
\nonumber   \omega_m &=& \omega_0 + m \omega_A
\end{eqnarray}
where $ E_m$ is the complex amplitude of the field component ${\cal E}_m(t)$. Equation \ref{Eq_cal_E_sum_over_harmonic} yields:
\begin{equation}\label{Eq_E_m}
    E_m = J_m(\Phi)~ E
\end{equation}

\begin{figure}
\begin{center}
\includegraphics[width=5.8 cm]{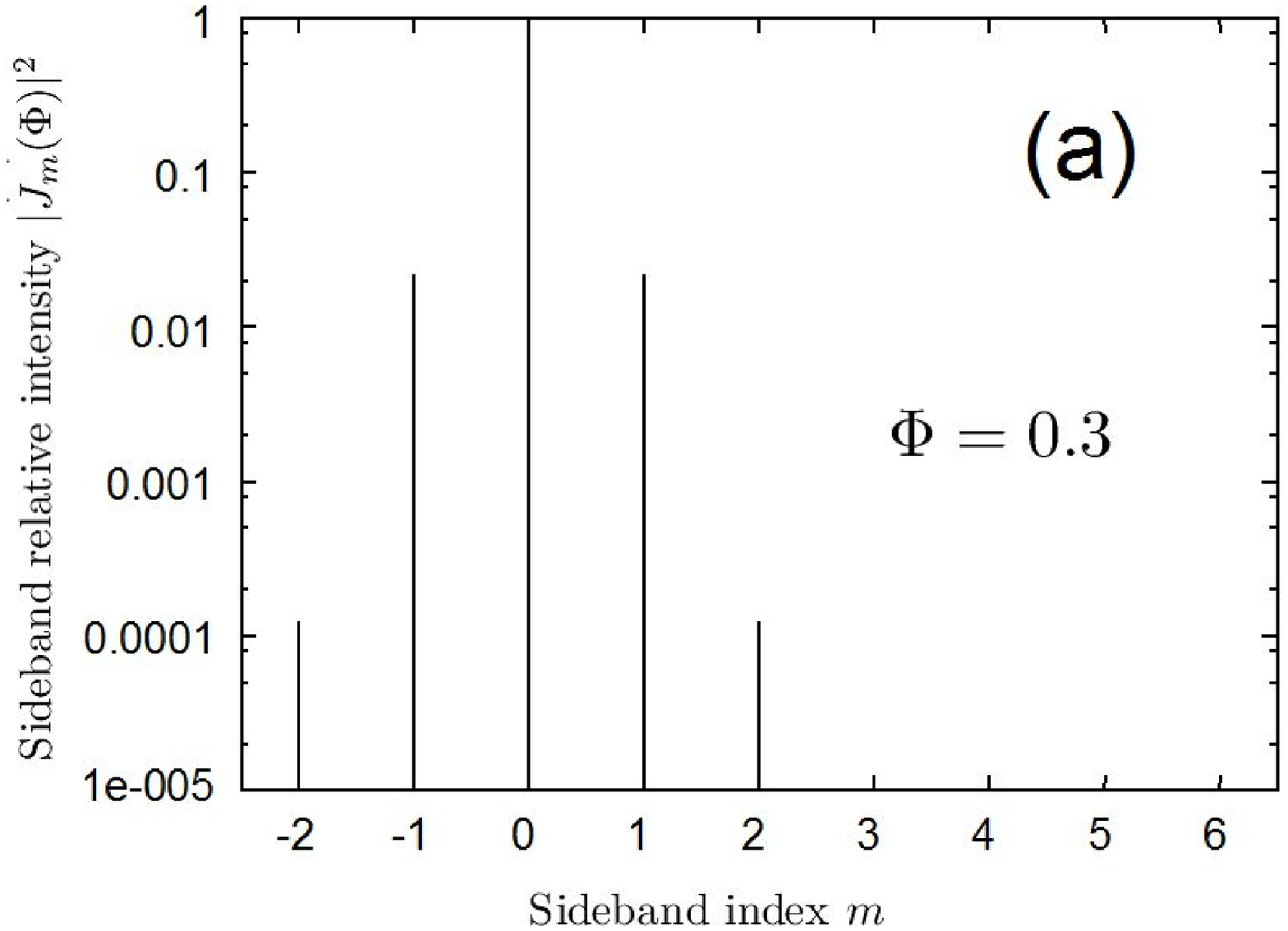}
\includegraphics[width=5.8 cm]{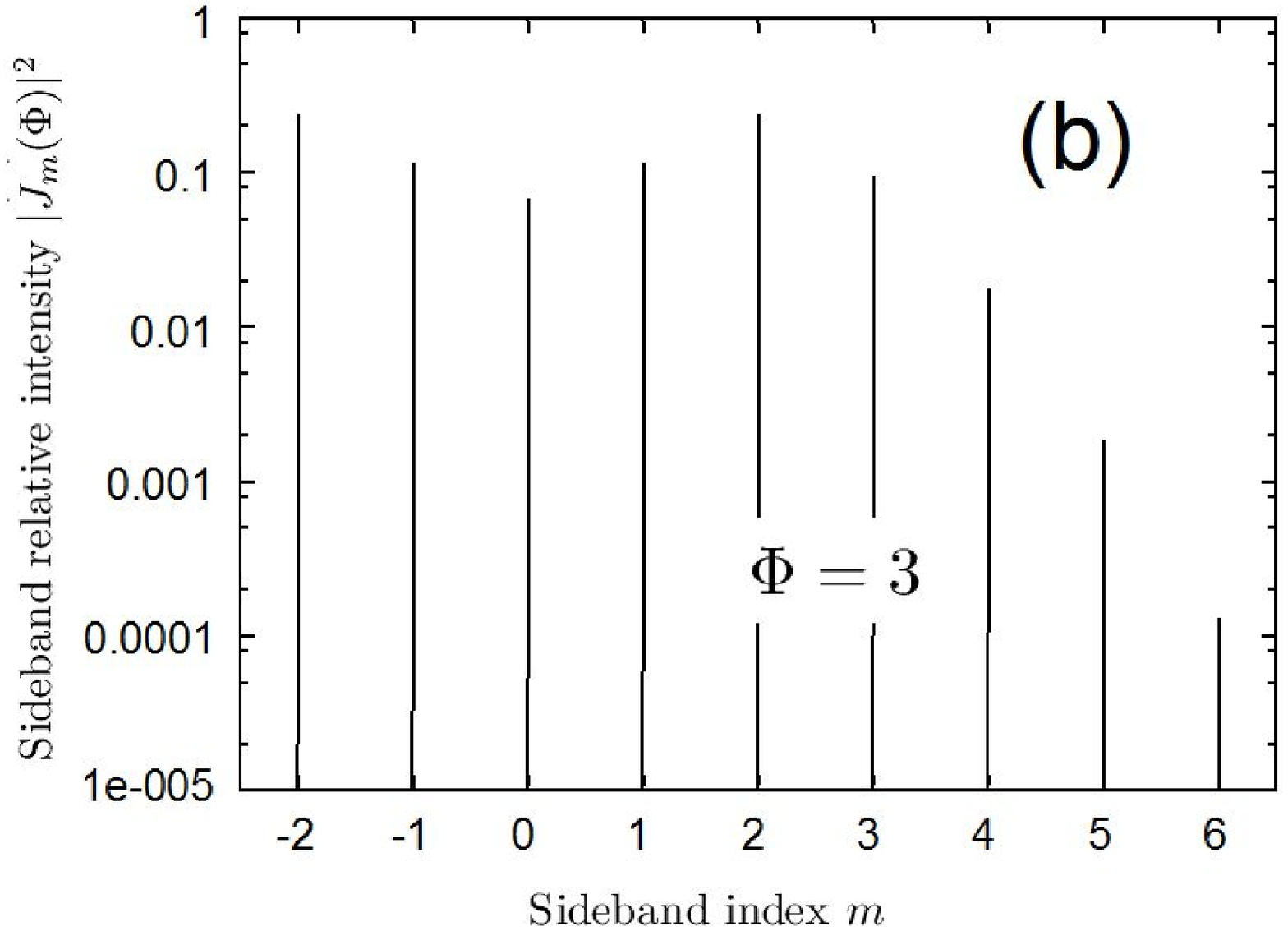}
 \caption{Relative amplitude $|E_m|^2/|E|2=|J_m(\Phi)|^2$ of the sideband component $m$ for an  amplitude modulation of the phase equal to  $\Phi=0.3$  (a) and $\Phi=3.0$ (b)  rad.}\label{Fig_fig_sideband _energy}
\end{center}
\end{figure}
Figure \ref{Fig_fig_sideband _energy} presents the distribution of the field energy   on the sidebands components $|E_m|^2$. If the amplitude of modulation $\Phi$ is low, most of the energy is on the carrier $|E_0|/|E|^2\simeq 1$, and energy $|E_m|^2$  decreases rapidly with the sideband index $m$. If the amplitude $\Phi$ is large, the energy of the carrier is low $|E_0|/|E|^2 \ll 1$, while  energy is distributed over many sidebands  $|E_m|^2$ .

\subsection{Selective detection of the sideband components $E_m$: sideband holography}
\label{section_Selective detection of the sideband components}

\begin{figure}
\begin{center}
\includegraphics[width=8 cm]{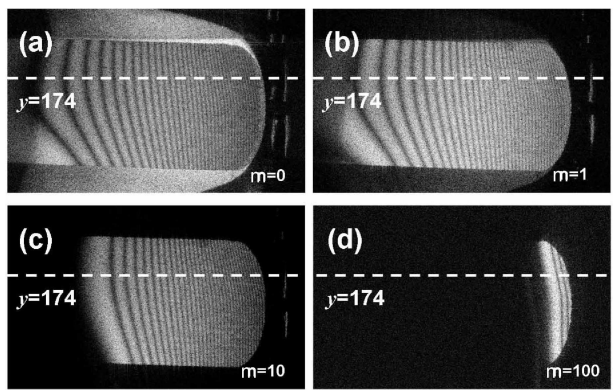}
 \caption{Reconstructed holographic images of a clarinet reed vibrating at frequency $\omega_A/2\pi =
2143$ Hz perpendicularly to the plane of the figure. Figure (a) shows the carrier image
obtained for $m = 0$. Fig. (b)-(d) show the frequency sideband images respectively for $m = 1$,
$m = 10$, and $m = 100$. A logarithmic grey scale has been used}\label{Fig_fig_clarinet_reed}
\end{center}
\end{figure}
Heterodyne holography that is able to perform
the holographic detection with any frequency shift $\omega-\omega_0 $ with respect to the
object illumination angular frequency $\omega_0$ is well suited to detect the vibration sideband components $E_m$.
To selectively detect  by  4 phase demodulation the sideband $m$ of frequency $\omega_m$,
the frequency $\omega_{LO}$ must be adjusted to fulfil the condition :
\begin{eqnarray}\label{Eq_Delata omega 4phase_sideband}
     \omega_{LO}&=& \omega_m-\omega_{CCD}/4\\
   \nonumber    &=& \omega_0 + m \omega_A -\omega_{CCD}/4
\end{eqnarray}

Figure \ref{Fig_fig_clarinet_reed} shows  images obtained by detecting different sideband $m$ of a clarinet reed \cite{joud2009imaging}.
The clarinet reed is
attached to a clarinet mouthpiece and its vibration is driven by a sound wave propagating inside
the mouthpiece, as in playing conditions, but the sound wave is created by a
loudspeaker excited at frequency $\omega_A$ and has a lower intensity than inside a clarinet. The excitation frequency
is adjusted to be resonant with the first flexion mode (2143 Hz) of the reed.

Figure \ref{Fig_fig_clarinet_reed} (a) is obtained at the unshifted carrier frequency $\omega_0$. It corresponds to an image obtained by time averaging holography \cite{picart2003time}.
The left side of the reed
is attached to the mouthpiece, and the amplitude of vibration is larger at the tip of the reed
on the right side; in this region the fringes become closer and closer and difficult to count.
The mouthpiece is also visible, but without fringes since it does not vibrate. Similar images
of clarinet reeds have been obtained in \cite{demoli2004detection,picart20052d}, with more conventional techniques and lower
image quality. Figures \ref{Fig_fig_clarinet_reed} (b) to (d) show images obtained for the sidebands $m=1$, 10 and 100. As
expected, the non-vibrating mouthpiece is no longer visible. Figure \ref{Fig_fig_clarinet_reed} (b) shows the $m = 1$
sideband image, with $J_1$ fringes that are slightly shifted with respect to those of $J_0$. Figure
\ref{Fig_fig_clarinet_reed} (c) shows the image of sideband  $m = 10$ and Fig.
\ref{Fig_fig_clarinet_reed} (d) the  $m = 100$ one. The left side region of the image
remains dark because, in that region, the vibration amplitude is not sufficient to generate these
sidebands, $J_m(z)$ being evanescent for $z < m$.

\begin{figure}
\begin{center}
\includegraphics[width=10 cm]{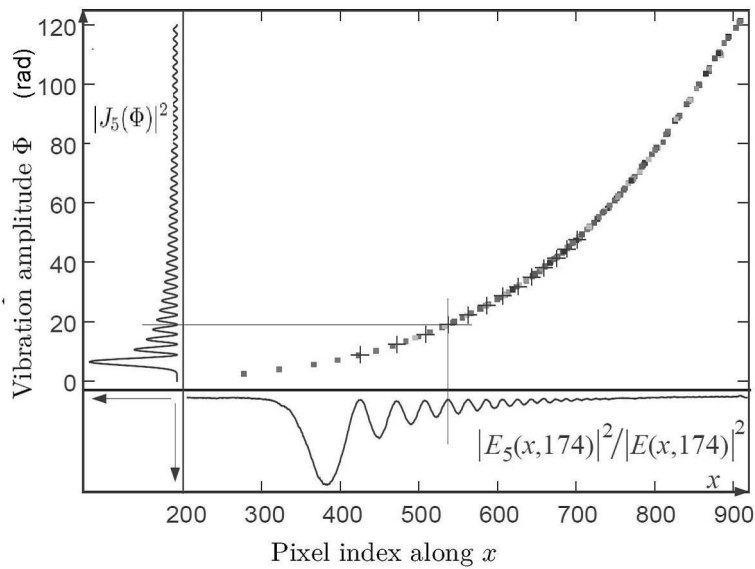}
 \caption{A slice of the data along the $y = 174$ line is used in this figure; the $x$ horizontal
axis gives the pixel index (100 pixels correspond to 3.68 mm.), the vertical axis the vibration
amplitude $\Phi$. The lower part of the figure shows the normalized signal
$|E_m(x)|^2/|E(x)|^2$  (where $E(x)$ is obtained loud-speaker off) for  $m = 5$, with a downwards axis. The left part
shows the corresponding square of the Bessel function $|J_5(\Phi)|^2 $ with a leftward axis. The
zeroes of the two curves are put in correspondence, which provides the points in the central
figure.   Similar correspondences are made harmonic order $m = 0, 1, 5, 10...100$.
Different gray densities are used for  different $m$.    The crosses
correspond to $m = 5$. The juxtaposition of the points for all values of $m$ gives an accurate
representation of the amplitude of vibration $\Phi$ as a function of $x$.}\label{Fig_fig_opex_curve}
\end{center}
\end{figure}

To quantitatively visualize the vibration amplitude $\Phi$, cuts of the reconstructed images signal $|E_m(x,y)|^2$ are made  for different
sideband orders $m$ along the horizontal line $y = 174$. This value has been chosen because it
corresponds to a region where the fringes are orthogonal to the $y$ axis.
To build the central part of Fig. \ref{Fig_fig_opex_curve},  the position of the antinodes of $|E_m(x,174)|^2$ are put in correspondence  with the antinodes of  $|J_m(\Phi)|^2 $. Correspondence is made  for  $m =$ 0, 1, 5, 10, . . . 100. Note that this  method is insensitive to inhomogeneities in the illumination zone. Therefore, no normalization is required. The curve seen in the central zone of Fig. \ref{Fig_fig_opex_curve} represents the amplitude of phase oscillation  $\Phi$ in radian  as a function of pixel index along $x$ for $y=174$.

\begin{figure}
\begin{center}
\includegraphics[width=10 cm]{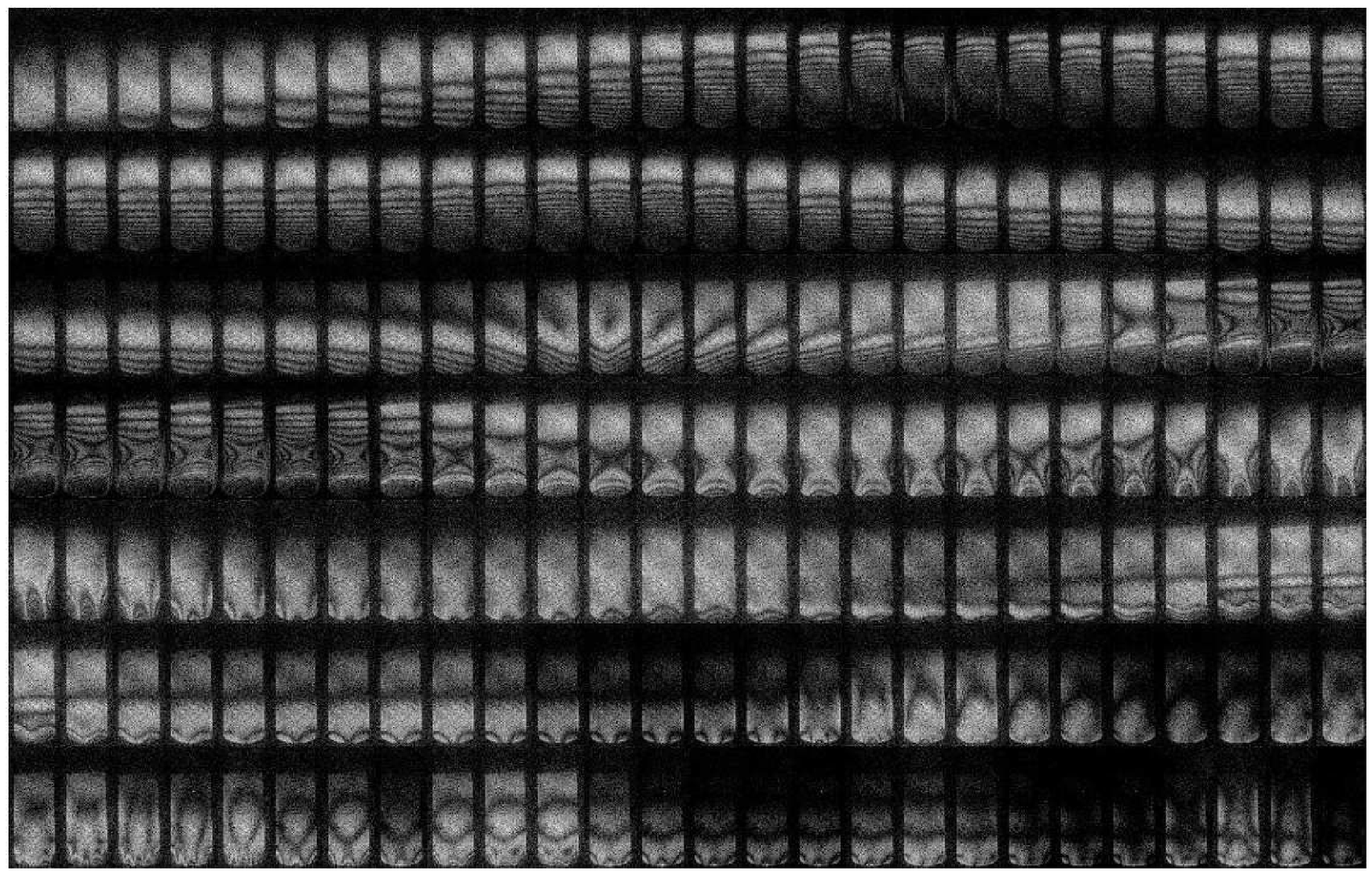}
 \caption{Clarinet reed  reconstructed images  obtained on sideband $m=1$.  Frequency $\omega_A$ is swept  from 1.4 KHz up to 20 KHz, and images are displayed from left to right and   top to bottom ($26 \times 7$ images). The display is made with arbitrary grey scale for the intensity $|E_0(x,y)|^2$. }\label{Fig_fig_taillard}
\end{center}
\end{figure}

\textbf{Remark:} In a typical heterodyne holography setup, digital synthesizers drive the acousto optic modulator at $\omega_{AOM1}$ and  $\omega_{AOM2}$,
the camera at  $\omega_{CCD}$, and the vibration frequency at $\omega_{A}$. These synthesizers use a common 10 MHz reference frequency, and  are driven by the computer.   It is then possible to automatically  sweep $\omega_{A}$, and $\omega_{AOM1}$  (or $\omega_{AOM2}$) in order to fulfil Eq. \ref{Eq_Delata omega 4phase_sideband} so that detection remains ever tuned on a given sideband.  Figure \ref{Fig_fig_taillard} shows an example \cite{taillard2014statistical}. A series of $26 \times 7$ images of a clarinet reed  are obtained on sideband $m=1$  by sweeping the frequency $\omega_A$ from 1.4 kHz up to 20 kHz by steps of 25 cents. The amplitude of the excitation signal is  exponentially increased in the range 1.4 to 4 kHz, from 0.5 to 16 V, then kept constant at 16 V up to 20 kHz. This crescendo limits the amplitude of vibration of the first two resonances of the reed.  The different vibration  modes  of the reed can be easily recognized on the reconstructed reed images of Fig. \ref{Fig_fig_taillard}.

\subsection{Sideband holography for large amplitude of vibration}\label{Section_large_vibration}

In the previous section (section \ref{section_Selective detection of the sideband components} and Fig. \ref{Fig_fig_opex_curve})
we have shown how the comparison of dark fringes for different sideband leads to a determination of the vibration
amplitude $\Phi(x, y)$ at each point of the object. This determination is non-local, since it involves counting fringes
from one reference point of the image to the point of interest, so that large amplitudes are not accessible.
The vibration amplitude $\Phi(x, y)$ can be determined by another approach that
completely eliminates the necessity of counting fringes,
giving a local measurement of the amplitude of vibration $\Phi$,
even for large $\Phi$ \cite{joud2009fringe}.

\begin{figure}
\begin{center}
\includegraphics[width=7 cm]{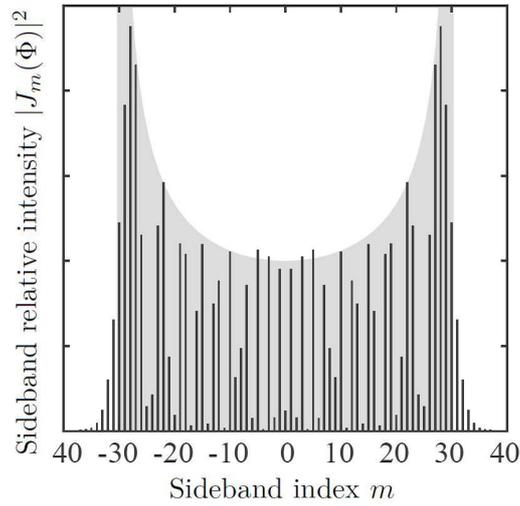}
 \caption{Relative intensity $|E_m|^2/|E|2=|J_m(\Phi)|^2$ of the sideband component $m$ for an  amplitude modulation of the phase equal to  $\Phi=30.3$  radiant. The light grey shade shows the Doppler spectrum obtained
from the vibration velocity distribution, with a continuous
variable $m$. Both spectra
fall abruptly beyond $m=30.3$, which corresponds to the
Doppler shift $\pm \omega_A \Phi$ associated with the maximum velocity.}\label{Fig_fig_Jm_303}
\end{center}
\end{figure}

For large amplitude of vibration ( $\Phi\gg 1$), the  distribution of the sideband energy $|E_m|^2$  over $m$    exhibits a sharp variation from maximum to zero near $m \simeq \Phi$, as seen on  Fig. \ref{Fig_fig_Jm_303} that plot $|E_m|^2/|E|^2$ for  $\Phi=30.3$. This property can be understood if one consider the limits $z_{max}\gg \lambda$. In that case, one can define an instantaneous  Doppler angular frequency shift $\omega_D(t)$, and an instantaneous sideband index $m(t)$ that are continuous variables :
\begin{eqnarray}
 \omega_D(t)&=&\omega_A \Phi \cos( \omega_A t) \\
 \nonumber m(t)&=& \frac{\omega_D(t)}{ \omega_A}=  \Phi \cos( \omega_A t)
\end{eqnarray}
Because of its sinusoidal variation, $m(t)$ spend more time near the extreme points $m= \pm \Phi$. The Doppler continuous distribution spectrum of $m(t)$ that is displayed in  light grey shade on Fig. \ref{Fig_fig_Jm_303} are thus maximum near the extreme  points $m= \pm \Phi$.

The vibration amplitude $\Phi(x,y)$ can be determined for each location $x,y$,  by measuring $|E_m(x,y)|^2$  for all sidebands $m$, and by determining for each location $x,y$ the sideband index $m$  that correspond to a fast drop of the signal  $|E_m(x,y)|^2$.    The method is robust and can easily be used even when the fringes become so narrow that they cannot be resolved, which gives immediate
access to large amplitudes of vibration.

\begin{figure}
\begin{center}
\includegraphics[width=10 cm]{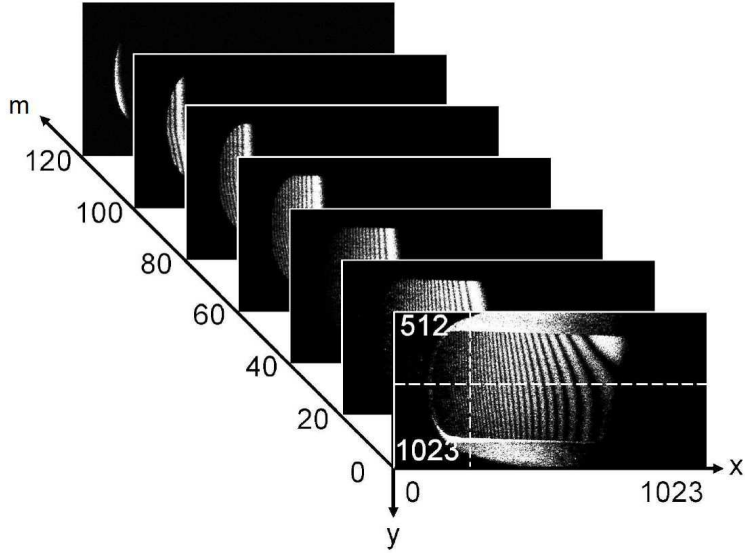}
 \caption{Cube of data obtained from the reconstructed holographic images of a vibrating clarinet reed. The sideband images
with $m = 0$, 20, 40 ...120 are shown in arbitrary linear scale.
 The white dashed lines correspond to $x = 249$
and $y = 750$, i.e. to the point chosen for Fig. \ref{Fig_fig3__OL}.}\label{Fig_fig_cube_of_data}
\end{center}
\end{figure}

By successively adjusting the frequency $\omega_{LO}$ of
the local oscillator to appropriate values, one records
the intensity images $|E_m(x,y)|^2$ of the sidebands as a
function of $x$ $y$ and $m$. One then obtains a cube of data
with three axes $x$, $y$ and $m$. Figure \ref{Fig_fig_cube_of_data} shows the images obtained for for $m = 0$, 20 ...120
that correspond to cuts of the cube along $x,y$ planes.
 The images illustrate how, when $m$ increases, the fringes move towards regions with larger amplitudes of vibrations. Since the right part of the reed ($x > 800$) is clamped on
the mouthpiece,  no signal is obtained in regions near the clamp
where $\Phi =4\pi z_{max}/\lambda \leq m$ (right part of the images).

\begin{figure}
\begin{center}
  \includegraphics[width=8 cm]{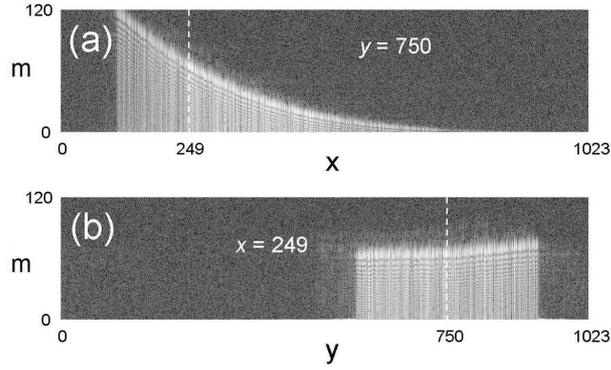}\\
  \caption{Images corresponding to cuts of 3D data of the reconstructed
 images along the planes $y=750$ (a) and $x=249$
(b). Fig~(a) shows the deformation of the object along its axis and
Fig~(b) a transverse cut with  a slight vibration asymmetry. A
logarithmic intensity scale is used.} \label{Fig_fig3__OL}
\end{center}
\end{figure}

Figure \ref{Fig_fig3__OL} (a) displays a 2D cut (coordinates $x,m$) of the cube of
data along the  horizontal plane $y = 750$ (horizontal
white dashed line in Fig. \ref{Fig_fig_cube_of_data}). The envelope of the non-zero
(non black) part of the image provides a measurement of
the amplitude of vibration in units of $\lambda/4\pi$. One actually
obtains a direct visualization of the shape of the reed at
maximal elongation, from the right part clamped on the
mouthpiece to the tip on the left. The maximum amplitude correspond to $\Phi=120$ rad.

\begin{figure}
\begin{center}
  \includegraphics[width=6 cm]{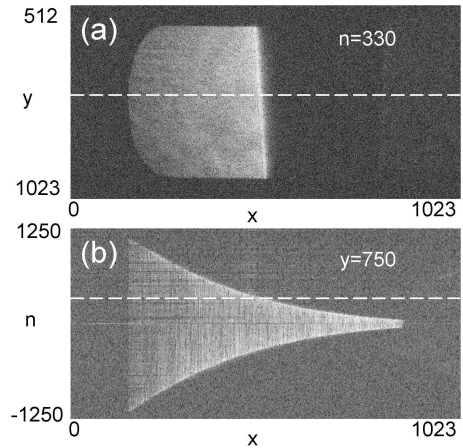}\\
 \caption{(a): image reconstructed with sideband $n=330$, with a large amplitude of vibration. (b) is the equivalent of
Fig.~\ref{Fig_fig3__OL}(a), but with positive  and negative $m$
values. One measures a maximum vibration amplitude of
$z_{ max} \simeq 60$ $\mu$m. A logarithmic intensity
scale is used.} \label{Fig_fig4__OL}
\end{center}
\end{figure}

Figure \ref{Fig_fig4__OL} shows  images
obtained at higher excitation amplitudes, about
10 times larger that for Fig.\ref{Fig_fig3__OL}. Figure \ref{Fig_fig4__OL} (a) shows the images
obtained for $m = 330$: the fringes are now completely
unresolved, but the transition from zero to non-zero intensity remains very clear. With a single hologram, and
without fringe counting, one obtains a clear marker of the line where $\Phi(x, y) = 330$ rad.
Figure \ref{Fig_fig4__OL}(b) shows the equivalent of Fig. \ref{Fig_fig3__OL}(a), but with
a higher excitation level, and this time for positive and
negative values of $m$.  Data range up to about $|m|\simeq 1140$,
corresponding to $z_{max}\simeq 58.4$ $\mu$m. Since the vibration
amplitude is much larger than $\lambda$, the continuous approximation for $m$ is valid, and the images of Fig. \ref{Fig_fig4__OL} can
be reinterpreted in term of classical Doppler effet.

Taking advantage of
the sideband order $m$  of the light scattered by a vibrating object adds a new dimension to digital holography.
Each pixel $x,y$ of the image then provides an information
that is completely independent from others, which results in redundancy and robustness of the measurements.
Looking at the edges of the spectrum provides an accurate determination of the vibration amplitude and avoids
a cumbersome analysis of the whole cube of data cube,
giving easy access to a measurements of large amplitudes
of vibration.

\subsection{Sideband holography with strobe illumination}\label{section_strobe_detection}

\begin{figure}[h]
\begin{center}
  \includegraphics[width=7 cm]{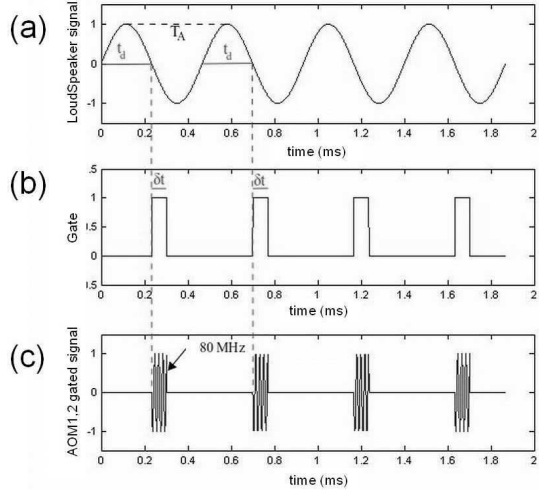}\\
 \caption{Chronogram of the signals.  Sinusoidal signal of period $T_A = 2\pi/\omega_A$  exciting
the reed (a). Rectangular gate that is applied to both illumination and reference beams (b) .Gate delay is $t_d$ with respect to the origin of reed sinusoidal motion. Gate duration is $\delta t$.  Gated AOM1 and AOM2 signals at $\simeq 80 $ MHz (c). These signals drive the acousto optic modulator  AOM1 and AOM2 and switch on and off the illumination and reference beams.} \label{Fig_fig_chronogramme}
\end{center}
\end{figure}

Both time averaged \cite{picart2003time} and sideband digital holography \cite{joud2009imaging} record the holographic signal over a large number a vibration periods. These two techniques are not sensitive to the phase of the vibration,
and are thus unable to measure the instantaneous velocities of the object. To respond this problem,
Leval et al. \cite{leval2005full} combine time averaged holography with stroboscopic illumination, but,
since Leval uses a mechanical stroboscopic device, the Leval technique suffer of a quite low
duty cycle (1/144), and is limited in low vibration frequencies ($\omega_A/2\pi < 5$ kHz).

\begin{figure}[h]
\begin{center}
  \includegraphics[width=5.0 cm]{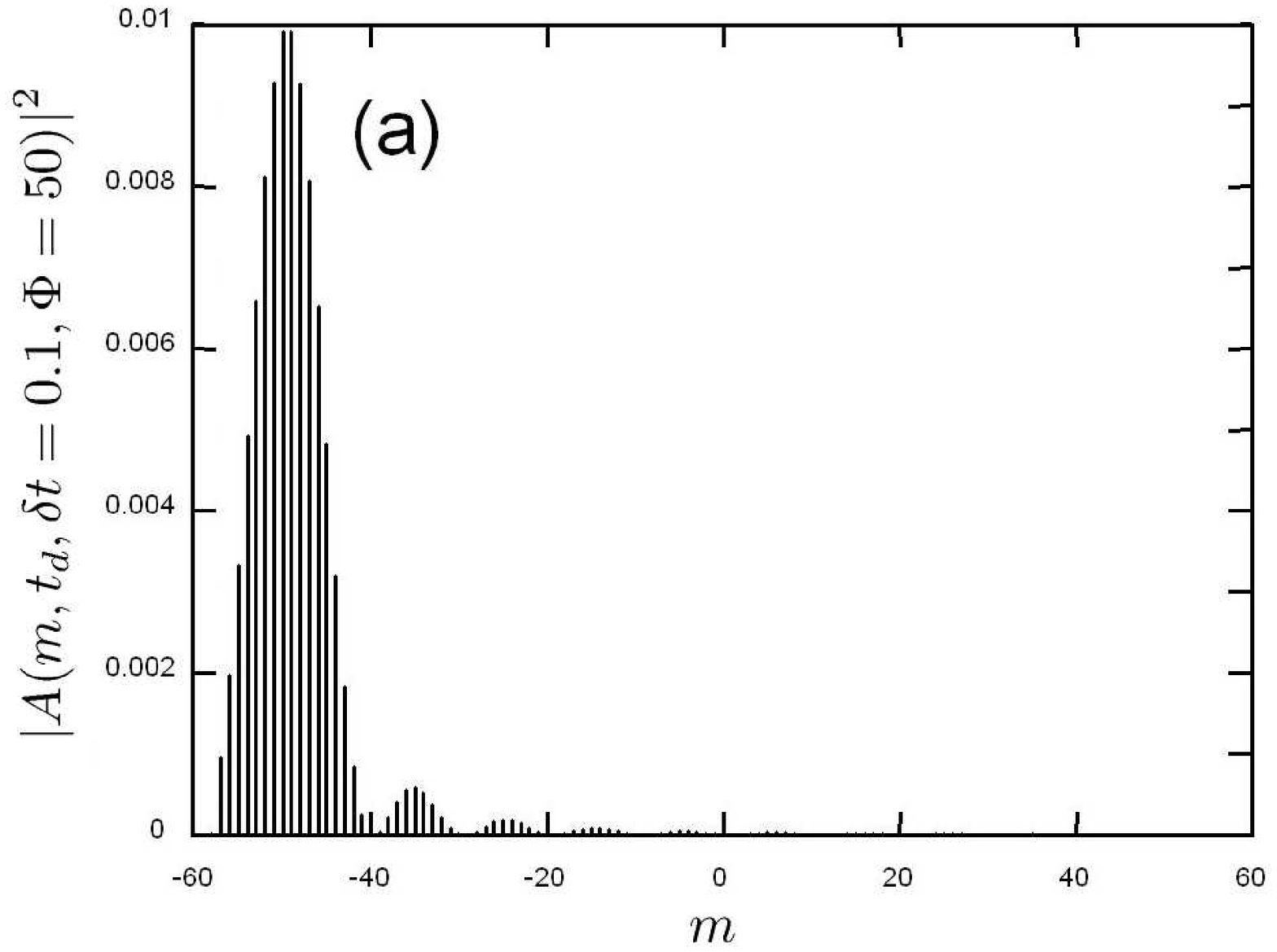}
  \includegraphics[width=5.0 cm]{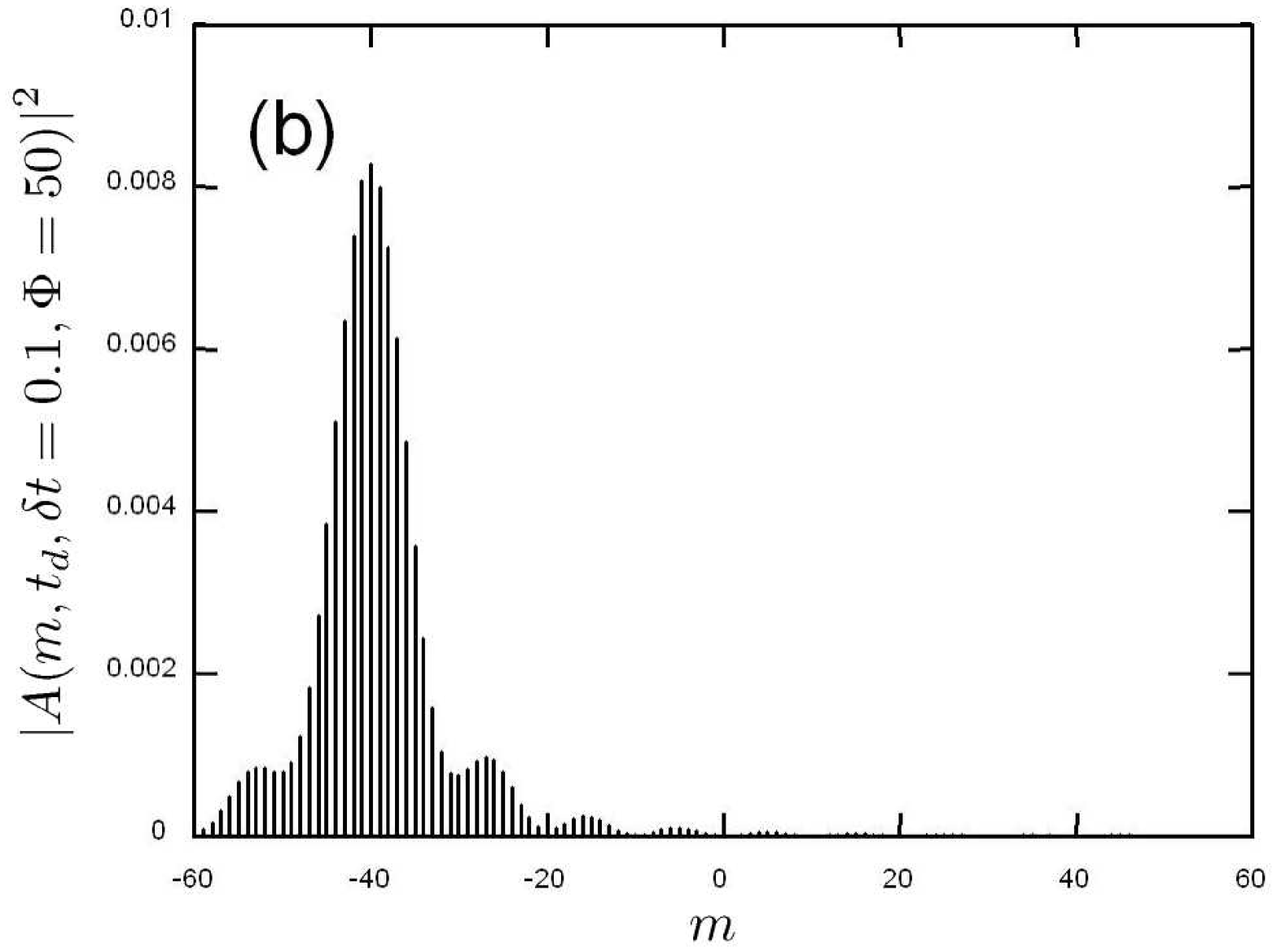}
  \includegraphics[width=5.0 cm]{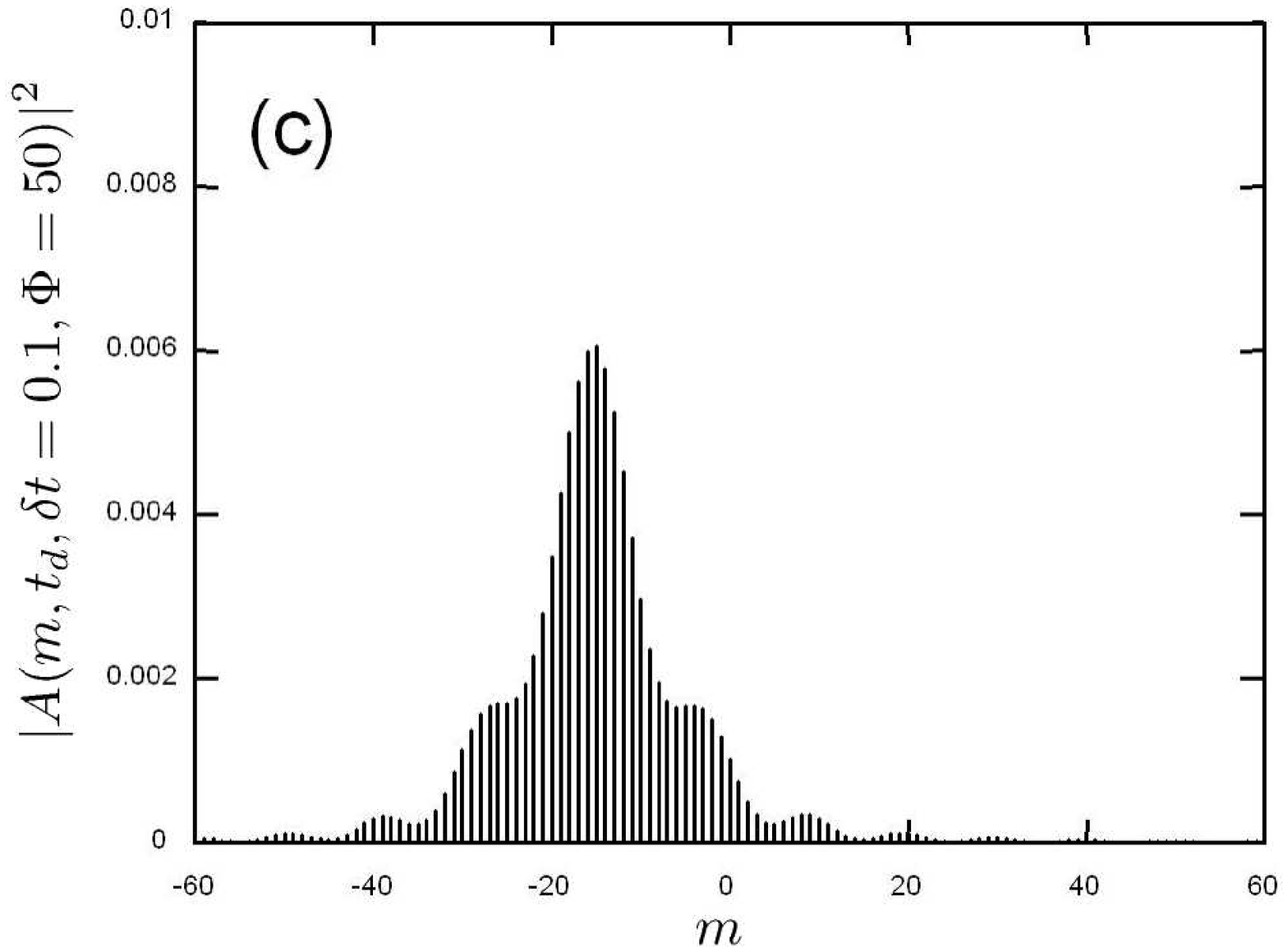}
  \includegraphics[width=5.0 cm]{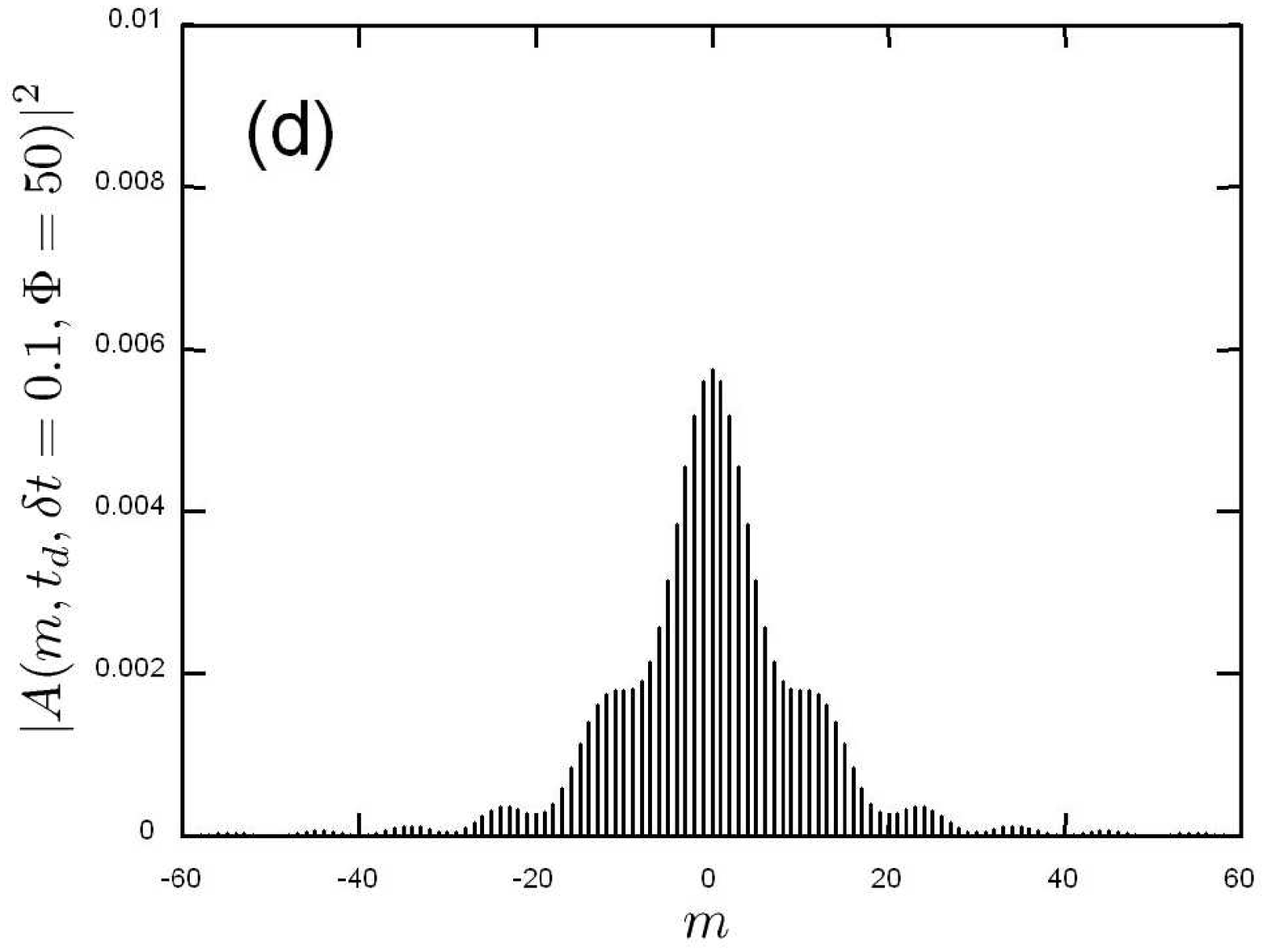}
 \caption{Doppler spectrum $|E_m|^2/|E|^2=|A(m,t_d,\delta t)|^2$ calculated for  a vibration amplitude  $\Phi=0.50$, a gate width $\delta t=0.1~ T_A$ and a gate time $t_c=0.25$ (a), 0.35 (b) , 0.45 (c) and 0.50 $T_A$ (d).} \label{Fig_fig_Am}
\end{center}
\end{figure}

To overcome these two limitations, it is possible to combine sideband digital holography  with stroboscopic illumination and detection
synchronized with the vibration \cite{verpillat2012imaging}. This can be achieved without changing the experimental sideband holography setup of Fig.\ref{Fig_fig_setup_reed} by switching electronically on and off the Radio Frequency signals that drive the AOMs at $\omega_{AOM1}$ and $\omega_{AOM2} \simeq 80 $ MHz. Figure \ref{Fig_fig_chronogramme} shows a typical chronogram of stroboscopic illumination and detection.

Without stroboscopic illumination (and detection), the Doppler  velocity spectrum is similar to the ones observed in Fig.\ref{Fig_fig_Jm_303} section \ref{Section_large_vibration}. It covers the entire range of speeds that can be reached during the sinusoidal motion, that is $\Phi<m<+\phi$, with two maxima near  $m \simeq \pm \Phi$. With stroboscopic illumination, the Doppler spectrum is modified, since it  corresponds to the narrower velocities range that is reached during gated illumination.
On the other hand, the finite gate duration   $\delta t$ (that is much shorter than the vibration period $T_A= 2\pi/\omega_A$) yields a broadening of  the spectrum of about  $T_A/\delta t$ for $m$.
In order to take quantitatively these effects into account, we have calculated the spectrum. The slowly varying complex field $E(t)$ must be multiplied by a gate
function $ H(t,t_d,\delta t)$ of period $T_A$. Within the interval $[0,T_A]$ we have:
\begin{eqnarray}\label{Eq_gate}
  H(t,t_d,\delta t) &=& 0 ~~~~\textrm{for}~~~~0<t<t_d-\delta t/2\\
\nonumber   &=&  1~~~~\textrm{for}~~~~t_d-\delta t/2<t<t_d+\delta t/2\\
 \nonumber  &=&0~~~~\textrm{for}~~~~t_d+\delta t/2<t<T_A
\end{eqnarray}
The slowly varying complex field signal $E(t)$ becomes thus:
\begin{eqnarray}\label{Eq_gated_signal}
E(t)&=& E~ H(t,t_d,\delta t)~ e^{j(\Phi \sin \omega_A t) }\\
\nonumber &=&E~\sum_m ~A(m,t_d,\delta t,\Phi)~e^{j m \omega_A t}\\
\nonumber E_m&=& E ~A(m,t_d,\delta t,\Phi)
\end{eqnarray}
where $A(m,t_d,\delta t)$ is the $m^{\textrm{th}}$ Fourier component of the periodic function $E(t)/E$.
We have calculated $A(m,t_d,\delta t,\Phi)$ by Fourier transformation of $E(t)/E$, and we have plotted on Fig. \ref{Fig_fig_Am} the Doppler spectrum  $|E_m|^2/|E|^2=|A(m,t_d,\delta t)|^2$ for  different gate  times $t_c$ and for $\delta t/T_A=0.1$. In Fig. \ref{Fig_fig_Am} (a), the gate  coincides with the maximum velocity ($t_d=0.5~T_A$). The Doppler spectrum is narrow and centered near $m\simeq  - \Phi = -50$ rad. In figure \ref{Fig_fig_Am} (b) and (c) i.e. for gate time $t_c=0.25$ (a), 0.35 (b), the absolute velocity decreases, and the center of  the Doppler spectrum moves towards $m=0$. On the other hand,  the Doppler spectrum broadens. In figure \ref{Fig_fig_Am} (c) the gate  coincides with the minimum  velocity ($t_d=0.5~T_A$). The center of the spectrum is $m=0$, and  the   broadening is maximum.

\begin{figure}
\begin{center}
  \includegraphics[width=11 cm]{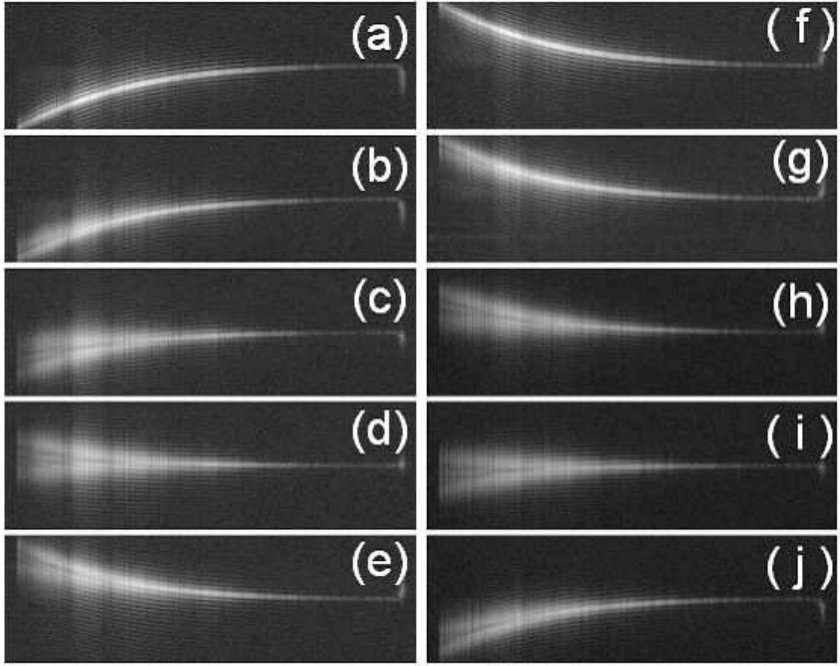}
 \caption{Successive velocities of the reed on a period $T_A$. These images are obtained by
making a cut  in the $y = 256$  plane  of the  3D stack of reconstructed images, whose coordinates are $x,y$ and $m$ with $m =$ -100 to +100. Gate duration is $\delta t=0.1 ~T_A$. From (a) to (j), the gate time $t_d$ is swept from $t_d=0.25 ~T_A$ to  $t_d=1.15~T_A$ by 0.1 $T_A$ step.   The images are displayed in logarithmic scale for the sideband complex field intensity: $|E_m(x,y=256)|^2$. } \label{Fig_anche___xz}
\end{center}
\end{figure}

To measure the instantaneous velocities of the object at time $t=t_d$, the gate is adjusted at $t_d$, and the holograms of all sidebands $m$ are recorded.  This can be done automatically  by the computer, by adjusting the frequency $\omega_{LO}$  to fulfil  Eq.\ref{Eq_Delata omega 4phase_sideband}. The sideband images  $E_m(x,y)$ are then reconstructed for all $m$, and  a 3D  cube of data (coordinate $x$, $y$ and $m$) is  obtained as illustrated by Fig. \ref{Fig_fig_cube_of_data}.
One can get velocity images by making cuts along $x$ or $y$. Figure \ref{Fig_anche___xz} shows an examples of  velocity images obtained at different times $t_d$ of the vibration period. Here, the reed is oriented in the same direction than in Fig.\ref{Fig_fig_cube_of_data}, and  the velocity images of coordinates $x,m$ are obtained  with horizontal cuts in plane $y=256$.

\subsection{Sideband holography for small amplitude of vibration}\label{section_small_vibration}

When the vibration amplitude $\Phi$ becomes small, the energy  within sidebands  decrease very rapidly with the sideband indexes $m$, and  one has only to consider the carrier $m=0$, and the two first sideband $m=\pm1$. Time averaged holography \cite{picart2003time} that detects the carrier field $E_0$  is  not efficient in detecting small amplitude vibration $\Phi$, because  $E_0$  varies quadratically  with $\Phi$:
\begin{eqnarray}\label{Eq_E_0_lim0}
E_0(\Phi) &=&  E~J_0(\Phi)  \\
 \nonumber  & \simeq & E~(1- \Phi^2)/6
\end{eqnarray}
%
On the other hand,  sideband holography that is able to detect the two  first sidebands fields $E_{\pm 1}$ is  much more sensitive, because  $E_{\pm 1}$  varies   linearly with $\Phi$:
\begin{eqnarray}\label{Eq_E_1_lim0}
  E_{\pm 1}(\Phi) &=& E~ J_{\pm 1}(\Phi)  \\
 \nonumber   &\simeq &\pm E ~\Phi/2
\end{eqnarray}
This point has been noticed about 40 years ago by Ueda et al. \cite{ueda1976signal}, who has made a  comprehensive study of the signal-to-noise
ratio (SNR) observed for classical (photographic film) sideband
holography. The authors managed to observe
vibration amplitudes of a few Angtroms, and linked
the smallest detectable amplitude to the SNR in the
absence of spurious effects. Later on, sub nanometric
vibration amplitude measurements were achieved with
sideband digital holography \cite{psota2012measurement,verrier2013absolute,bruno2014phase,bruno2014holographic,bruno2014non}, and comparison with  single point  laser interferometry has been  made \cite{psota2012comparison,bruno2014holographic}.

One must notice that the complex field amplitude $E$ scattered by the sample without vibration  depends strongly on the $x$ $y$ position. In a typical experiment, the sample rugosity is such that  $E$ is a speckle that is fully developed. The field $E(x,y)$ is thus random in amplitude and phase from one speckle to the next. One cannot thus  extract the phase of the vibration motion from a measurement made on single a sideband, for instance the sideband $m=1$. To get the phase, one needs to measure the field components on several sidebands, for example  the carrier field $E_0$  and the first sideband $E_1$. For small amplitude $\Phi$, the carrier field $E_0(\Phi)$ is very close to the field $E=E_0(\Phi=0)$ measured without vibration.   By measuring both $E_0(\Phi)$ and $E_1(\Phi)$ by sideband digital holography, one can then eliminate $E$ in Eq. \ref{Eq_E_0_lim0} and Eq. \ref{Eq_E_1_lim0}
getting by the way $J_0(\Phi)$, $J_1(\Phi)$  and $\Phi$. This can be made by calculating  either the ratio $E_1/E_0$ \cite{verrier2013absolute}: %
\begin{eqnarray}\label{Eq_E1_E0_ratio}
  \left|\frac{E_1}{E_0}\right| &=& \frac{J_1(\Phi)}{J_0(\Phi)}\\
\nonumber  &\simeq& \Phi/2
\end{eqnarray}
or the correlation $E_1~E^*_0$:
\begin{eqnarray}\label{Eq_E1_E0_correlation}
\left| E_1~E^*_0 \right| &=& |E|^2~ J_1(\Phi) ~J_0(\Phi)\\
\nonumber &\simeq & |E|^2 \Phi/2
\end{eqnarray}
Both methods yield a signal ($\Phi/2$ or $|E|^2 \Phi/2 $) that is proportional to $\Phi$.

\begin{figure}
\begin{center}
  \includegraphics[width=11 cm]{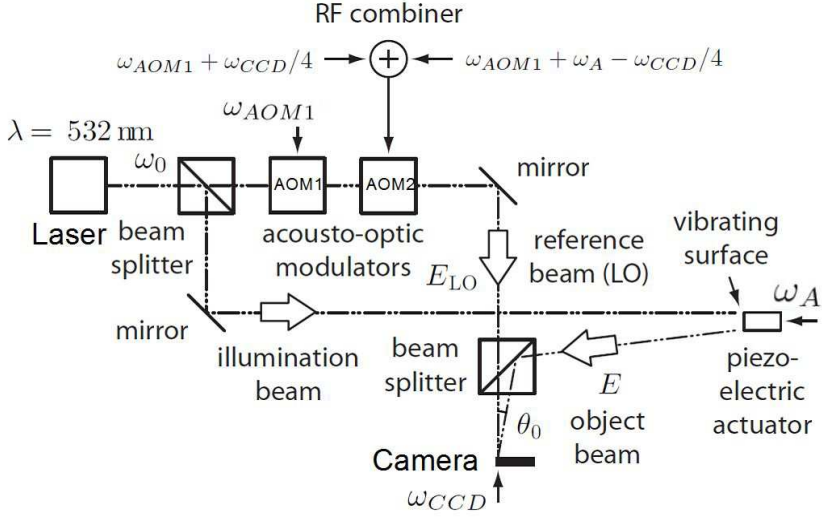}
 \caption{Sideband holography setup able to acquire the carrier and sideband signal simultaneously} \label{Fig_fig_2_sideband_setup}
\end{center}
\end{figure}

From a practical point of view,   $E_0$ and $E_1$ are measured successively in a standard sideband holography setup like the one of  Fig.\ref{Fig_fig_setup_reed} \cite{joud2009imaging,psota2012measurement}.  A sequence of $n_{max}$ frames with $\omega_{LO}=\omega_{0}- \omega_{CCD}/4$ is first recorded yielding  $E_0$. A second sequence  with $\omega_{LO}=\omega_{0}+\omega_A -\omega_{CCD}/4$ is then recorded yielding  $E_1$.

It is also possible to record $E_0$ and $E_1$ simultaneously \cite{verrier2013absolute,bruno2014phase,bruno2014holographic,bruno2014non}. In that case, the RF signal that drives the second acousto optic modulator AOM2  is made of two  frequency components (Fig. \ref{Fig_fig_2_sideband_setup}) at $\omega_{AOM2}^a$ and  $ \omega_{AOM2}^b$:
\begin{eqnarray}
  \omega_{AOM2}^a &=& \omega_{AOM1}+\omega_{CCD}/4 \\
\nonumber \omega_{AOM2}^b &=& \omega_{AOM1}+\omega_A-\omega_{CCD}/4
\end{eqnarray}
\begin{itemize}
   \item The first component at $\omega_{AOM2}^a$, whose weight is  $\alpha$,  yields the carrier signal $E_0$, if the hologram $H$ is calculated with demodulation on the -1 grating order i.e. with $ H=\sum_{n=0}^{n_{max}} (-j)^n I_n $.
   \item The second component at $\omega_{AOM2}^a$, whose weight is  $\beta$, yields the sideband $E_1$ with demodulation on the +1 grating order i.e. with $ H=\sum_{n=0}^{n_{max}} (+j)^n I_n $.
 \end{itemize}

\begin{figure}
\begin{center}
  \includegraphics[width=11 cm]{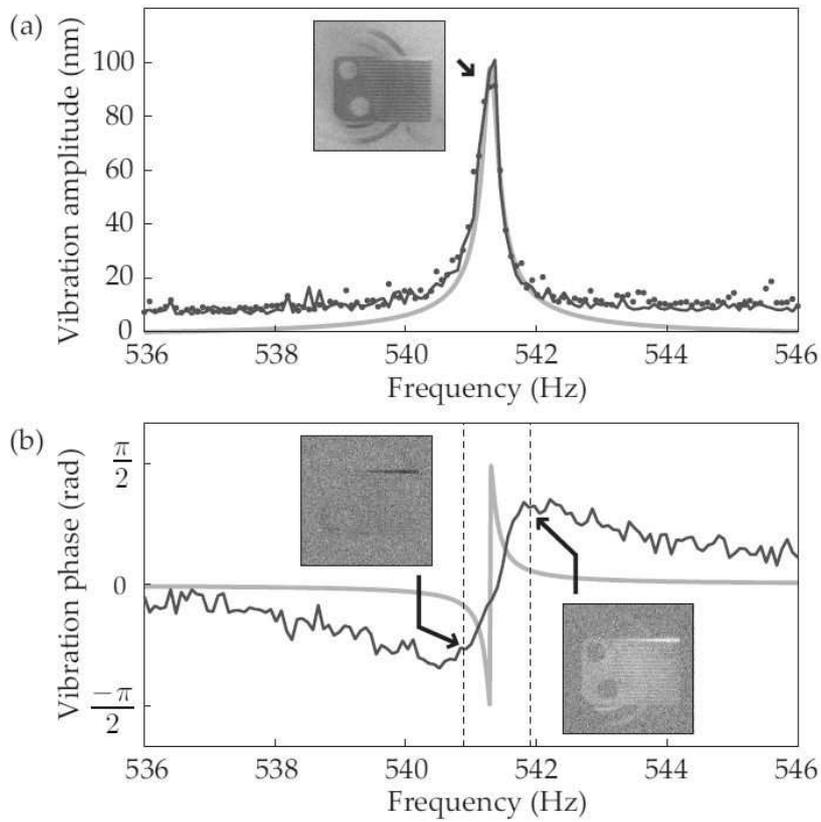}
 \caption{Vibration amplitude $z_{max}$ (a) and phase $\varphi$ (b) of a lamellophone, averaged over the first upper cantilever,  versus excitation frequency $\omega_A/(2\pi)$ (Hz). Insets: retrieved vibration amplitude and phase maps in the neighborhood of the resonance. The theoretical resonance line is in gray.} \label{Fig_fig_canteviller_mike}
\end{center}
\end{figure}

Simultaneous detection of $E_0$ and $E_1$, and calculation of $\Phi$ by the ratio method of Eq. \ref{Eq_E1_E0_ratio} has been used to study various vibrating samples. Figure \ref{Fig_fig_canteviller_mike} shows example of signals obtained with a lamellophone of a musical box \cite{bruno2014phase}. In that experiment, all the frequencies $\omega_{A}$, $\omega_{CCD}$, $\omega_{AOM1}$, $\omega_{AOM2}^a$ and  $\omega_{AOM2}^b$  are driven by numerical synthesizers. In this particular case, harmonics are extracted by a temporal Fourier transform of a sequence of 8 raw interferograms. Here, optical sidebands of interest are downconverted to demodulated frequencies in the camera bandwidth. In order to avoid signal phase drifts from the measurement of one sequence to the next, AOM2 is driven using a phase ramp. As a consequence, the phases of $E_0$ and $E_1$ are well defined, and  yield the phase $\varphi$ of the mechanical motion. We have:
\begin{eqnarray}\label{equ_phase_mechnical}
  \varphi &=&\textrm{Arg} (E_1/E_0) +C
\end{eqnarray}
where $C$ is a additive constant that depends of the relative phase of the synthesizers. Figure \ref{Fig_fig_canteviller_mike} shows  the averaged  amplitude $z_{max}$ and phase $\varphi$ of the upper cantilever of the lamellophone, which is driven though its resonance frequency $\simeq 541.5$ Hz. As expected, the phase $\varphi$ makes a jump of about $\pi$ when excitation frequency $\omega_A$ crosses the resonance frequency.

\begin{figure}
\begin{center}
  \includegraphics[width=11 cm]{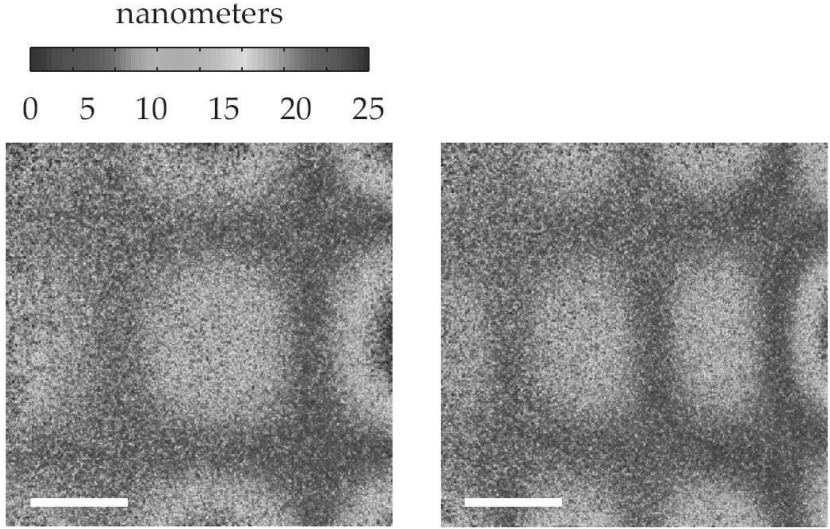} \caption{Amplitude maps of the out-of-plane vibration of a thin metal plate versus excitation frequency $\omega_A/(2\pi)$. Holographic
images at 40.1 kHz (a), 61.7 kHz (b). Scalebar: 5 mm.} \label{Fig_fig_acoustic_wave}
\end{center}
\end{figure}

\begin{figure}
\begin{center}
  \includegraphics[width=11 cm]{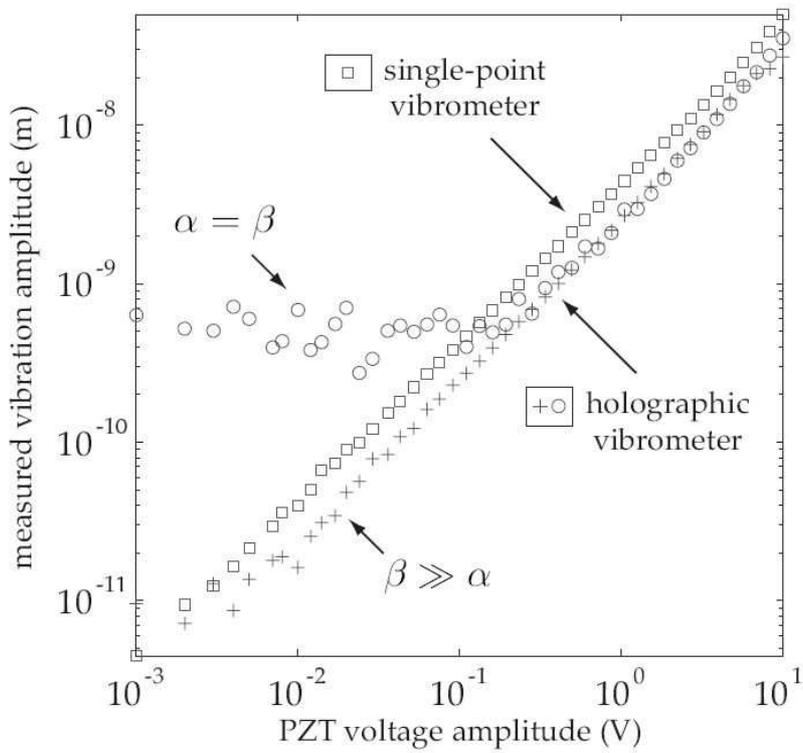}
 \caption{ Comparison of quantitative out-of-plane vibration
amplitudes $z_{max} $ retrieved with the single-point laser vibrometer ($ \square$
symbols) and holographic vibrometer with $\beta=\alpha$ ($\circ$ symbols)
and $\beta\simeq 50 \alpha$ ($+$ symbols).} \label{Fig_curve_holo_versus_laser_doppler}
\end{center}
\end{figure}

The method has been also applied to get full field  images of  surface acoustic  waves as shown on Fig. \ref{Fig_fig_acoustic_wave} \cite{bruno2014holographic}. Comparison with single point Doppler vibrometry is reported in Fig.~\ref{Fig_curve_holo_versus_laser_doppler}. If the carrier and sideband components of the RF signal at $\omega_{AOM2}^a$ and  $\omega_{AOM2}^b$ have the same amplitude (i.e. if $\alpha=\beta$), some spurious detection of the  carrier field $E_0$ (which is much bigger) is observed  when the detection is tuned on the grating order -1 (used to detect  $E_0$). This spurious effect limits the detection sensitivity to $z_{max}\sim 1 $ nm, as seen on Fig. \ref{Fig_fig_acoustic_wave} ($\circ$ symbols). Much better results are obtained by reducing the weight of the carrier component (i.e. with $\beta\gg \alpha$). The detection sensitivity becomes  $z_{max}\sim 0.01 $ nm with $\beta\simeq 50 \alpha$  as seen on Fig. \ref{Fig_fig_acoustic_wave} ($+$ symbols).

The ratio method of Eq. \ref{Eq_E1_E0_ratio} used above is simple, since it gives directly  $\Phi$. Nevertheless, the method is unstable for the points $x,y$ of the object where $E$ is close to zero.  Since $\Phi$ varies slowly with the location $x,y$, one can improve the accuracy by averaging  $\Phi$ over neighbor points. In that case, the ratio method is not optimal for noise, since all points  $x,y$ are weighted equally in the average, while the accuracy on the ratio depends strongly on the location  $x,y$, since the accuracy is low when  $|E (x,y)|$ is low.

For this issue, the correlation method of Eq.\ref{Eq_E1_E0_correlation} is less simple, since it gives $|E|^2\Phi$, and not $\Phi$. It is nevertheless better for noise, since spatial averaging over neighbor points $x,y$ is weighted by the scattered energy $|E(x,y)|^2$.

\begin{figure}
\begin{center}
  \includegraphics[width=11 cm]{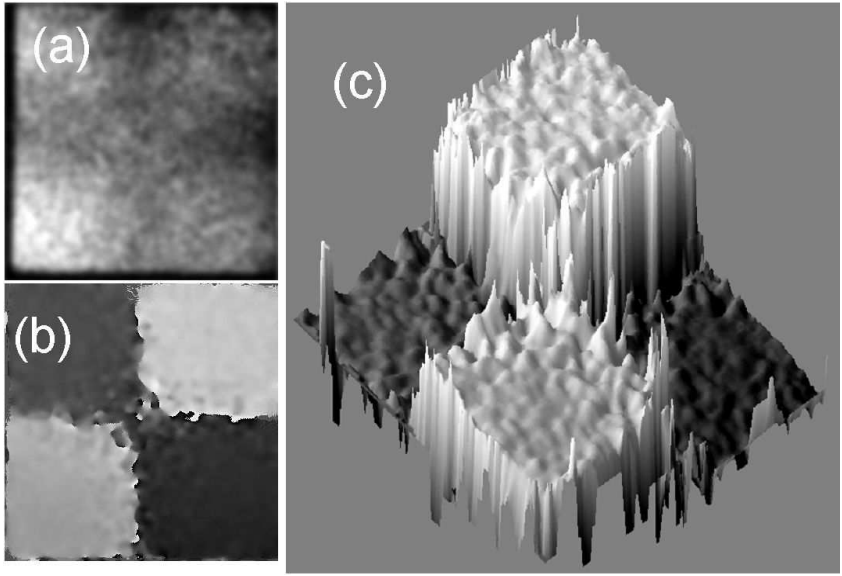}
 \caption{(a,b) Reconstructed images of the correlation $E_1 E^*_0$ obtained with a cube of wood; (a) amplitude: $|E_1 E^*_0|$ ,  and (b) phase: $\arg E_1 E^*_0$. The correlation $E_1(x,y) E^*_0(x,y)$ is averaged with a 2D gaussian blur filter of radius 4 pixels. The display is made in arbitrary linear scale. (c) 3D display of  the phase: $x,y,\arg E_1 E^*_0$} \label{Fig_fig_vibrating_cube}
\end{center}
\end{figure}

Figure \ref{Fig_fig_vibrating_cube} shows reconstructed images of the correlation $E_1 E^*_0$ obtained with a cube of wood (2 cm $\times$ 2 cm) vibrating at its resonance frequency $\omega_A=21.43 $ kHz. Here, the fields $E_0$ and $E_1$ are measured sucessively.  In order to get better SNR, the complex correlation signal $E_1(x,y) E^*_0(x,y)$ is averaged over neighbor $x,y$ points by using a 2D gaussian blur filter of radius 4 pixels. Figure \ref{Fig_fig_vibrating_cube}  shows the amplitude (a) and phase (b) of the the filtered correlation $E_1(x,y) E^*_0(x,y)$. Figure \ref{Fig_fig_vibrating_cube} (c) displays the phase of  the correlation   in 3D. As seen,  the opposite corners (upper left and bottom right for example) vibrate in phase, while the neighbor corners (upper left and upper right for example) vibrate in phase opposition. Note that the cube is excited in one of its corner by a needle. This may explain why the opposite corners are not perfectly in phase in Fig. \ref{Fig_fig_vibrating_cube} (b,c).

\begin{figure}
\begin{center}
  \includegraphics[width=9 cm]{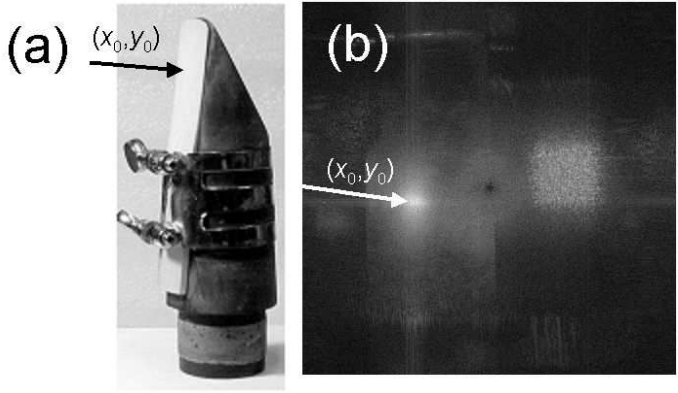}
 \caption{ (a) Clarinet reed with illumination beam focused in $x_0,y_0$. (b) Sideband $m=1$ reconstructed image of the vibrating reed. The display is made in arbitrary log scale for the field intensity $|E_1(x,y)|^2$.   } \label{Fig_fig_reed_1pt}
\end{center}
\end{figure}

In order to increase the ability to analyze vibration of small amplitude, one can  increase the illumination on the zone where the correlation averaging is performed. This is  done by using a collimated illumination beam. Figure \ref{Fig_fig_reed_1pt} shows an example. The illumination laser has been focused on point $x_0, y_0$ of  the clarinet read to be studied (see Fig. \ref{Fig_fig_reed_1pt} (a) ). Most of the  energy in the reconstructed image is thus located near $x_0,y_0$ (see Fig. \ref{Fig_fig_reed_1pt} (b) ). To measure the vibration amplitude at location $x_0,y_0$, we have calculated the averaged correlation  $\langle E_1 E^*_0 \rangle $ and the averaged energy $\langle |E_0|^2 \rangle $:
\begin{eqnarray} \label{Eq_averaged E0E1}
  \langle E_1 E^*_0 \rangle &=& (1/N_{pix}) \sum_{x,y} E_1(x,y) E^*_0(x,y) \\
  \nonumber  &=&  J_1(\Phi)  J_0(\Phi)~ (1/N_{pix})\sum_{x,y} |E(x,y)|^2
\end{eqnarray}
\begin{eqnarray} \label{Eq_averaged E0E0}
\langle |E_0|^2 \rangle &=&(1/N_{pix}) \sum_{x,y} |E_0(x,y)|^2 \\
  \nonumber  &=&    J^2_0(\Phi)~ (1/N_{pix})\sum_{x,y} |E(x,y)|^2
\end{eqnarray}
where $\sum_{x,y}$ is the summation over the $N_{pix}$ pixels of the illuminated region  located near $x_0,y_0$. In that region, the vibration amplitude $\Phi$ is supposed to  not depend on $x$ and $y$. We get then:
\begin{eqnarray} \label{Eq_Phi_E0EO_E1E0}
   \frac{ \langle E_1 E^*_0 \rangle   }{ \langle |E_0|^2 \rangle }=\frac{J_1(\Phi)}{J_0(\Phi)} \simeq \Phi/2
\end{eqnarray}

\begin{figure}
\begin{center}
  \includegraphics[width=9 cm]{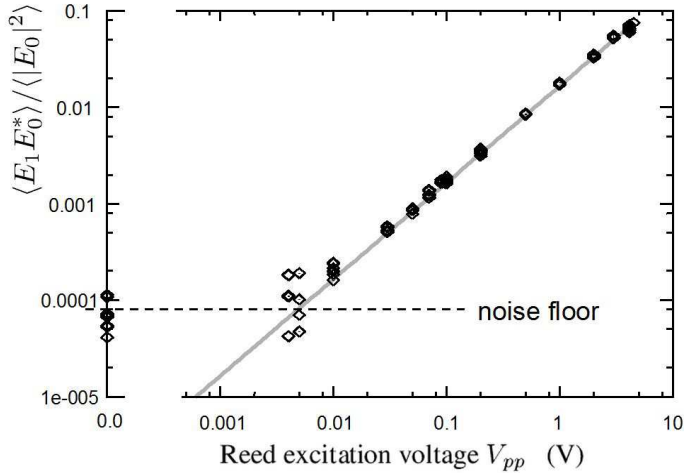}
 \caption{ Ratio $ \langle E_1 E^*_0 \rangle/ \langle |E_0|^2 \rangle$  (vertical axis) as function of the reed excitation voltage $V_{pp}$ in volt Units.  The light grey curve  is $J_1(\Phi)/J_0(\Phi)$ with $\Phi = z_{max}/(4 \pi \lambda)$.} \label{Fig_fig_lobna}
\end{center}
\end{figure}


To verify the ability of measuring small vibration, the reed has been excited with lower and lower loudspeaker levels, the illumination laser being focused on the point $x_0,y_0$ of Fig. \ref{Fig_fig_reed_1pt} (b), we want to study. A Sequence of $n_{max}=128$ frames have been  recorded for both carrier ($E_0$) and sideband ($E_1$). Four phase detection has been then  made by calculating $H$  with $n_{nmax}=128$ in Eq. \ref{EQ_H_n_max}. We have then  calculated the reconstructed fields $E_0(x,y)$ and $E_1(x,y)$, the averaged correlation $\langle E_1 E^*_0 \rangle$ and averaged carrier intensity $\langle |E_0|^2 \rangle$ and the ratio $\langle E_1 E^*_0 \rangle  /\langle |E_0|^2 \rangle$. This ratio is plotted on Fig.\ref{Fig_fig_lobna} ($y$ axis)  as a function of the peak to peak voltage $V_{pp}$ of the sinusoidal signal that excites the loudspeaker  ($x$ axis). The measured ratio varies linearly with $V_{pp}$. For low voltage $V_{pp}\leq 5$ mV, the ratio reached a noise floor corresponding to $\langle E_1 E^*_0 \rangle  /\langle |E_0|^2 \rangle \simeq 10^{-4}$. Since this voltage $V_{pp}$ is proportional to the vibration amplitude $z_{max}$ and thus to $\Phi$, we have convert the  voltage $V_{pp}$ in vibration amplitude  $z_{max}$. The conversion factor ($z_{max}=0.02$ nm for $V_{pp}=10$ mV) is obtained by assuming that  $J_1(\Phi)/J_0(\Phi)$ fits the ratio $\langle E_1 E^*_0 \rangle  /\langle |E_0|^2 \rangle$ that is  measured, as predicted by Eq. \ref{Eq_Phi_E0EO_E1E0} (see solid grey curve of Fig.\ref{Fig_fig_lobna}). We get here a noise floor noise corresponding to $z_{max}\simeq 0.01 $ nm. It   is about $\times 10$ lower than the limit $\lambda/3500 =0.22$ nm predicted by Ueda \cite{ueda1976signal}, and  $\times 10$ lower than in  previous experiments at similar  frequency \cite{psota2012measurement,verrier2013absolute} in the kHz range. Similar noise floor $0.01$ nm has been  obtained  by the ratio method on Fig. \ref{Fig_curve_holo_versus_laser_doppler}, but at higher vibration frequency $\omega \simeq 40 $ kHz  \cite{bruno2014holographic}.

We have tried to make the experiment with more frames i.e. with $n_{max}>128$, but this does not lower the noise floor. The remaining noise floor can  be related  to a spurious detection of the carrier field $E_0$ when detection is tuned to detect $E_1$, or to some technical noise on the carrier signal that is not totally filter off.  Since the carrier and sideband fields $E_0$ and $E_1$ are within the same spatial mode, no space filtering can be applied to filter off $E_0$.  Here, the spurious carrier field $E_0$ is filtered off in the time domain, while the local oscillator field $E_{LO}$ is filtered off both in time and frequency domain.

\section{Conclusion}

In this chapter we have presented the digital heterodyne holography technique that is able to fully control the amplitude, phase and frequency of both illumination and reference beams. Full automatic data acquisition of the holographic signal can be made, and   ultimate shot noise sensitivity can be reached. Heterodyne holography is an extremely versatile and powerful tool, in particular when applied to vibration analysis. In that case, heterodyne holography is able to detect in wide field (i.e. in all points of the object 2D surface at the same time) the vibrating object signal at any optical sideband of rank $m$. Since the control of the intensity of the illumination and reference beams is fast, instantaneous measurements of the vibration signal, sensitive to the mechanical phase, can be made. For the measurement of large vibration amplitudes, the possible ambiguity of measurements  can be removed by making measurements at different sideband indexes $m$. For small  vibration amplitudes, the mechanical phase can be obtained from measurements made on the carrier ($m=0$) and on the first sideband ($m=1$). Moreover, extremely low vibration amplitudes, below 10 picometers, can be measured.

\bibliographystyle{unsrt}

\begin{thebibliography}{}

\end{thebibliography}


\begin{thebibliography}{10}

\bibitem{schnars1994direct}
Ulf Schnars and Werner Juptner.
\newblock Direct recording of holograms by a ccd target and numerical
  reconstruction.
\newblock {\em Applied optics}, 33(2):179--181, 1994.

\bibitem{yamaguchi1997phase}
Ichirou Yamaguchi and Tong Zhang.
\newblock Phase-shifting digital holography.
\newblock {\em Optics letters}, 22(16):1268--1270, 1997.

\bibitem{pedrini1995digital}
G~Pedrini, YL~Zou, and HJ~Tiziani.
\newblock Digital double-pulsed holographic interferometry for vibration
  analysis.
\newblock {\em Journal of Modern Optics}, 42(2):367--374, 1995.

\bibitem{pedrini1997digital}
Giancarlo Pedrini, Hans~J Tiziani, and Yunlu Zou.
\newblock Digital double pulse-tv-holography.
\newblock {\em Optics and lasers in Engineering}, 26(2):199--219, 1997.

\bibitem{pedrini1998transient}
G~Pedrini, Ph~Froening, H~Fessler, and HJ~Tiziani.
\newblock Transient vibration measurements using multi-pulse digital
  holography.
\newblock {\em Optics \& Laser Technology}, 29(8):505--511, 1998.

\bibitem{pedrini2006high}
Giancarlo Pedrini, Wolfgang Osten, and Mikhail~E Gusev.
\newblock High-speed digital holographic interferometry for vibration
  measurement.
\newblock {\em Applied optics}, 45(15):3456--3462, 2006.

\bibitem{fu2007vibration}
Yu~Fu, Giancarlo Pedrini, and Wolfgang Osten.
\newblock Vibration measurement by temporal fourier analyses of a digital
  hologram sequence.
\newblock {\em Applied optics}, 46(23):5719--5727, 2007.

\bibitem{powell1965interferometric}
Robert~L Powell and Karl~A Stetson.
\newblock Interferometric vibration analysis by wavefront reconstruction.
\newblock {\em JOSA}, 55(12):1593--1597, 1965.

\bibitem{picart2003time}
Pascal Picart, Julien Leval, Denis Mounier, and Samuel Gougeon.
\newblock Time-averaged digital holography.
\newblock {\em Optics letters}, 28(20):1900--1902, 2003.

\bibitem{picart2005some}
Pascal Picart, Julien Leval, Denis Mounier, and Samuel Gougeon.
\newblock Some opportunities for vibration analysis with time averaging in
  digital fresnel holography.
\newblock {\em Applied optics}, 44(3):337--343, 2005.

\bibitem{le2000numerical}
Fr{\'e}d{\'e}rique Le~Clerc, Laurent Collot, and Michel Gross.
\newblock Numerical heterodyne holography with two-dimensional photodetector
  arrays.
\newblock {\em Optics letters}, 25(10):716--718, 2000.

\bibitem{le2001synthetic}
Fr{\'e}d{\'e}rique Le~Clerc, Michel Gross, and Laurent Collot.
\newblock Synthetic-aperture experiment in the visible with on-axis digital
  heterodyne holography.
\newblock {\em Optics Letters}, 26(20):1550--1552, 2001.

\bibitem{atlan2007accurate}
Michael Atlan, Michel Gross, and Emilie Absil.
\newblock Accurate phase-shifting digital interferometry.
\newblock {\em Optics letters}, 32(11):1456--1458, 2007.

\bibitem{verpillat2010digital}
Fr{\'e}d{\'e}ric Verpillat, Fadwa Joud, Michael Atlan, and Michel Gross.
\newblock Digital holography at shot noise level.
\newblock {\em Journal of Display Technology}, 6(10):455--464, 2010.

\bibitem{gross2007digital}
Michel Gross and Michael Atlan.
\newblock Digital holography with ultimate sensitivity.
\newblock {\em Optics letters}, 32(8):909--911, 2007.

\bibitem{gross2008noise}
Michel Gross, Michael Atlan, and Emilie Absil.
\newblock Noise and aliases in off-axis and phase-shifting holography.
\newblock {\em Applied optics}, 47(11):1757--1766, 2008.

\bibitem{lesaffre2012noise}
Max Lesaffre, Nicolas Verrier, and Michel Gross.
\newblock Noise and signal scaling factors in digital holography in week
  illumination: relationship with shot noise.
\newblock {\em Appl. Opt.}, 52:A81--A91, 2013.

\bibitem{cuche2000spatial}
Etienne Cuche, Pierre Marquet, and Christian Depeursinge.
\newblock Spatial filtering for zero-order and twin-image elimination in
  digital off-axis holography.
\newblock {\em Applied Optics}, 39(23):4070--4075, 2000.

\bibitem{joud2009imaging}
Fadwa Joud, F~Lalo{\H{U}}, Michael Atlan, Jean Hare, and Michel Gross.
\newblock Imaging a vibrating object by sideband digital holography.
\newblock {\em Optics express}, 17(4):2774--2779, 2009.

\bibitem{demoli2004detection}
Nazif Demoli and Dalibor Vukicevic.
\newblock Detection of hidden stationary deformations of vibrating surfaces by
  use of time-averaged digital holographic interferometry.
\newblock {\em Optics letters}, 29(20):2423--2425, 2004.

\bibitem{picart20052d}
Pascal Picart, Julien Leval, Jean~Claude Pascal, Jean~Pierre Boileau, Michel
  Grill, Jean~Marc Breteau, Benjamin Gautier, and St{\'e}phane Gillet.
\newblock 2d full field vibration analysis with multiplexed digital holograms.
\newblock {\em Optics express}, 13(22):8882--8892, 2005.

\bibitem{taillard2014statistical}
Pierre-Andr{\'e} Taillard, Franck Lalo{\"e}, Michel Gross, Jean-Pierre Dalmont,
  and Jean Kergomard.
\newblock Statistical estimation of mechanical parameters of clarinet reeds
  using experimental and numerical approaches.
\newblock {\em Acta Acustica united with Acustica}, 100(3):555--573, 2014.

\bibitem{joud2009fringe}
Fadwa Joud, Fr{\'e}d{\'e}ric Verpillat, Franck Lalo{\"e}, M~Atlan, Jean Hare,
  and Michel Gross.
\newblock Fringe-free holographic measurements of large-amplitude vibrations.
\newblock {\em Optics letters}, 34(23):3698--3700, 2009.

\bibitem{leval2005full}
Julien Leval, Pascal Picart, Jean~Pierre Boileau, and Jean~Claude Pascal.
\newblock Full-field vibrometry with digital fresnel holography.
\newblock {\em Applied optics}, 44(27):5763--5772, 2005.

\bibitem{verpillat2012imaging}
Fr{\'e}d{\'e}ric Verpillat, Fadwa Joud, Michael Atlan, and Michel Gross.
\newblock Imaging velocities of a vibrating object by stroboscopic sideband
  holography.
\newblock {\em Optics express}, 20(20):22860--22871, 2012.

\bibitem{ueda1976signal}
Mitsuhiro Ueda, Sumio Miida, and Takuso Sato.
\newblock Signal-to-noise ratio and smallest detectable vibration amplitude in
  frequency-translated holography: an analysis.
\newblock {\em Applied optics}, 15(11):2690--2694, 1976.

\bibitem{psota2012measurement}
Pavel Psota, Vit Ledl, Roman Dolecek, Jiri Erhart, and Vaclav Kopecky.
\newblock Measurement of piezoelectric transformer vibrations by digital
  holography.
\newblock {\em Ultrasonics, Ferroelectrics and Frequency Control, IEEE
  Transactions on}, 59(9):1962--1968, 2012.

\bibitem{verrier2013absolute}
Nicolas Verrier and Michael Atlan.
\newblock Absolute measurement of small-amplitude vibrations by time-averaged
  heterodyne holography with a dual local oscillator.
\newblock {\em Optics letters}, 38(5):739--741, 2013.

\bibitem{bruno2014phase}
Francois Bruno, Jean-Baptiste Laudereau, Max Lesaffre, Nicolas Verrier, and
  Michael Atlan.
\newblock Phase-sensitive narrowband heterodyne holography.
\newblock {\em Applied optics}, 53(7):1252--1257, 2014.

\bibitem{bruno2014holographic}
Francois Bruno, J\'{e}r\^{o}me Laurent, Daniel Royer, and Michael Atlan.
\newblock Holographic imaging of surface acoustic waves.
\newblock {\em Applied Physics Letters}, 104(8):083504, 2014.

\bibitem{bruno2014non}
Francois Bruno, J{\'e}r{\^o}me Laurent, Claire Prada, Benjamin Lamboul, Bruno
  Passilly, and Michael Atlan.
\newblock Non-destructive testing of composite plates by holographic
  vibrometry.
\newblock {\em Journal of Applied Physics}, 115(15):154503, 2014.

\bibitem{psota2012comparison}
Pavel Psota, V{\'\i}t L{\'e}dl, Roman Dole{\v{c}}ek, Jan V{\'a}clav{\'\i}k, and
  Miroslav {\v{S}}ulc.
\newblock Comparison of digital holographic method for very small amplitudes
  measurement with single point laser interferometer and laser doppler
  vibrometer.
\newblock In {\em Digital Holography and Three-Dimensional Imaging}, pages
  DSu5B--3. Optical Society of America, 2012.

\end{thebibliography}

\end{document}